\documentclass[a4paper,10pt]{article}

\usepackage[margin=1in]{geometry}

\usepackage[english]{babel}
\usepackage{amssymb}
\usepackage{amsmath}
\usepackage{amsthm}
\usepackage{psfrag}
\usepackage[T1]{fontenc}
\usepackage{ae,aecompl}
\usepackage[colorlinks]{hyperref}
\usepackage{subfigure}
\usepackage{appendix}
\usepackage{natbib}
\hypersetup{colorlinks,breaklinks=true,citecolor=blue,linkcolor=blue}
\usepackage{graphicx}
\usepackage{color}

\usepackage{etoolbox}
\BeforeBeginEnvironment{tabular}{\begin{center}\small}
\AfterEndEnvironment{tabular}{\end{center}}

\usepackage{caption}
\providecommand\bnabla{\boldsymbol{\nabla}}

\newcommand\eg{e.g.\ }
\newcommand\ie{i.e.\ }

\newcommand{\pdt}[1]{\ensuremath{\frac{\mbox{$\partial$} #1}{\mbox{$\partial$} t}}}
\newcommand{\pl}{\left(}
\newcommand{\pr}{\right)}
\newcommand{\n}{\nabla}

\newcommand{\vect}[1]{\boldsymbol{#1}}
\newcommand{\vel}{\mathbf{u}}

\newcommand{\vorz}{\zeta}
\newcommand{\moyp}[1]{\overline{#1}}
\newcommand{\moyz}[1]{\left \langle #1 \right \rangle}

\newcommand{\zw}{\overline{u_{\phi}}}
\newcommand{\urms}{\mbox{\textit{Re}}}
\newcommand{\uzrms}{\mbox{\textit{Re}}_0}
\newcommand{\ucrms}{\mbox{\textit{Re}}_c}
\newcommand{\Roz}{\mbox{\textit{Ro}}_0}
\newcommand{\Roc}{\mbox{\textit{Ro}}_c}
\newcommand{\us}{u_s^{\ast}}
\newcommand{\PV}{q}

\newcommand{\Pran}{\mbox{\textit{Pr}}} 
\newcommand{\Ek}{\mbox{\textit{Ek}}}
\newcommand{\Ra}{\mbox{\textit{Ra}}}
\newcommand{\Ro}{\mbox{\textit{Ro}}}

\newcommand{\Rey}{\mbox{\textit{Re}}}

\newcommand{\revision}[1]{#1}

\title{Multiple zonal jets and convective heat transport barriers in a quasi-geostrophic model of planetary cores}
\author{\large C\'eline Guervilly$^1$ \& Philippe Cardin$^{2}$  \vspace{0.2cm} \\
  {\small $^1$School of Mathematics, Statistics and Physics, Newcastle University, Newcastle upon Tyne, NE17RU, UK} \\
  {\small $^2$Institut des Sciences de la Terre, Universit\'e Grenoble Alpes, CNRS, 38041 Grenoble, France}
  }

\begin{document}

\maketitle


\begin{abstract}

We study rapidly-rotating Boussinesq convection driven by internal heating in a full sphere. We use a numerical model based on the quasi-geostrophic approximation for the velocity field, whereas the temperature field is three-dimensional.
This approximation allows us to perform simulations for Ekman numbers down to $10^{-8}$, Prandtl numbers relevant for liquid metals ($\sim10^{-1}$) and Reynolds numbers up to $3\times10^4$.
Persistent zonal flows composed of multiple jets form as a result of the mixing of potential vorticity.
For the largest Rayleigh numbers computed, the zonal velocity is larger than the convective velocity despite the presence of boundary friction.
The convective structures and the zonal jets widen when the thermal forcing increases.
Prograde and retrograde zonal jets are dynamically different: in the prograde jets (which correspond to weak potential vorticity gradients) the convection transports heat efficiently
and the mean temperature tends to be homogenised; by contrast, in the cores of the retrograde jets (which correspond to steep gradients
of potential vorticity) the dynamics is dominated by the propagation of Rossby waves, resulting in the formation of steep mean temperature gradients
and the dominance of conduction in the heat transfer process.
Consequently, in quasi-geostrophic systems, 
the width of the retrograde zonal jets controls the efficiency of the heat transfer.
\end{abstract}


\section{Introduction}

Convection is the main heat transport process in the liquid cores of planets and is thought to be responsible for the generation of planetary magnetic fields. 
Convection is strongly affected by the rapid rotation of the planet via the action of the Coriolis force. 
Owing to the very low fluid viscosity, the convective flows are turbulent, although the nonlinear inertial effects are relatively weak compared with the Coriolis force.
Under these conditions, and in the absence of magnetic fields, the primary dynamical balance is established between the Coriolis force and the pressure gradient
and is called geostrophic balance. Geostrophic flows are invariant along the rotation axis, and so,
in spherical geometry, they can only be axisymmetric and azimuthal (\ie zonal). 
Convective flows, which are directed along the direction of gravity, cannot be exactly geostrophic, but nevertheless form tall columnar flows aligned with the rotation axis \citep[][]{Jon15}; \revision{such flows are commonly referred to as ``quasi-geostrophic''}.
These columnar convective flows produce coherent Reynolds stresses that drive geostrophic zonal flows \citep{Gil77,Bus82}.
Stress-free boundary conditions, where boundary friction is absent, favour the emergence of strong zonal flows \citep[\eg][]{Aur01}. 
In models with relatively small viscosity (which can be measured by the Ekman number, the ratio of the rotation period to the
global viscous timescale), the zonal flows develop persistent multiple jets of alternating sign  \revision{inside the tangent cylinder} \citep[\eg][]{Hei05,Gastine2014}.
The observation of intense jets in geophysical and astrophysical objects \citep[\eg][]{Schou1998,Por03,Livermore2017} has prompted 
much effort dedicated to their study, and in particular, their width and amplitude \citep[\eg][]{Chr02,Gil07,Read2015,Cabanes2017}.
Although zonal flows (and shear flows in general) do not transport heat \revision{outwards}, they strongly affect the convection because they can deflect and shear the convective flows, 
thereby reducing the efficiency of the heat transfer \citep[\eg][]{Aurnou2008,Goluskin2014,Hardenberg2015,Yad16}.
In the present paper, we explore the effect of intense, multiple zonal jets on the convective heat transport in turbulent rotating convection for small Ekman numbers.

The numerical modelling of turbulent rotating flows is extremely challenging as it necessitates a wide range of dynamical length and time scales.
Numerical models must therefore employ Ekman numbers that are several orders of magnitude larger than those found in natural objects.
However, in the absence of magnetic fields, the lengthscale of the convective flows scales with the Ekman number, at least at the linear onset of convection. 
The coherence of the Reynolds stresses, and hence the width and amplitude of the zonal flows, might well be affected by the convective lengthscale, 
and thus by the Ekman number. 
In order to approach turbulent rotationally-constrained convection at small Ekman numbers, 
we alleviate part of the computational limitations by using a quasi-geostrophic (QG) 
approximation that was developed
by \citet{Bus86} for thermal convection in the annulus geometry of \citet{Bus70} with curved boundaries. The model neglects the variations of the axial vorticity of the flow
along the rotation axis, which allows to compute the velocity in two dimensions (2D). This is an important limitation to the full dynamics of rotating convection
\revision{\citep[\eg][]{Calkins2013}}, 
but the rationale of using this QG model is that it allows the exploration of currently inaccessible regions of the parameter space, thereby informing future three-dimensional studies.
Variations of the QG model have been successfully applied in numerous studies in spherical geometry \citep[\eg][]{Car94,Mor04,Cal12}.
Where possible, results from these studies have been successfully benchmarked against asymptotic theories \citep{Gil06,Lab15}, three-dimensional (3D) numerical models \citep{Aub03,Pla08}, and
laboratory experiments \citep{Aub03, Sch05,Gil07}.

Following the model constructed in \citet{Guervilly2016}, we use a hybrid numerical model that couples the QG velocity to a 3D implementation 
of the temperature in the whole sphere, in order to account for the spherical symmetry of the basic temperature background. 
\revision{The buoyancy driving is controlled by the temperature averaged along the direction of the rotation axis, which,
contrary to QG models using a 2D temperature field \citep{Bus86}, is not assumed to be equal to the temperature in the equatorial plane. 
Solving the temperature in 3D will allow us to assess the influence of the 3D temperature on the quasi-geostrophic dynamics.}
This implementation is particularly appropriate to model fluids with small Prandtl numbers (the ratio of the viscosity to the thermal diffusivity) that are typical
of liquid metals ($\mathcal{O}(10^{-1})$) by permitting the use of a 3D grid for the temperature that is coarser than the 2D grid used for the velocity.
For simplicity, we consider only thermal convection in a full sphere without a solid inner core. The thermal convection is driven by an homogenous internal heating, which 
is more relevant for the early history of the Earth's core.

The existence of a so-called strong branch of convection driven by internal heating, \revision{as first suggested by the weakly nonlinear analysis of \citet{Sow77},}
was recently found numerically by \citet{Guervilly2016} with the hybrid QG-3D model and by \citet{Kaplan2017} with a fully 3D model for Ekman numbers smaller than $\mathcal{O}(10^{-7})$ and Prandtl numbers smaller than unity.
The bifurcation is subcritical at the onset of convection and the strong branch is characterised by Reynolds numbers greater than $1000$ near the onset
and strong zonal flows. 
In this paper, we focus on the production of zonal flows on this strong branch of convection for $\Ek\in[10^{-8},10^{-7}]$ and $\Pran\in[10^{-2},10^{-1}]$.

The layout of the paper is as follows.
In \S\ref{sec:model}, we detail the formulation of the hybrid QG-3D model. In \S\ref{sec:structure}, we describe the radial dependence of the convection and zonal flows
and quantify the dependence of the jet width, convective lengthscale and zonal flow velocity on the model parameters. The drift and stability of the zonal flows is discussed in \S\ref{sec:stability} and their mechanism of formation in \S\ref{sec:PV}. The effect of the zonal flows on the heat transport is presented in \S\ref{sec:barrier}. Finally, a discussion
of the results is given in \S\ref{sec:discussion}.

\section{Mathematical formulation}
\label{sec:model}

We study Boussinesq thermal convection driven by internal heating in a rotating sphere.
The rotation vector is \mbox{$\Omega \vect{e}_z$}, where $\Omega$ is constant.
The acceleration due to gravity is radial and linear, \mbox{$\vect{g}=g_0 r \vect{e}_r$}.
The radius of the sphere is $r_o$ and no inner core is present.
The fluid has kinematic viscosity $\nu$, thermal diffusivity $\kappa$, density $\rho$, heat capacity 
at constant pressure $C_p$, and thermal expansion coefficient $\alpha$, all of which are constant.
We consider an homogeneous internal volumetric heating $S$. 
In the absence of convection, the static temperature profile $T_s$ is calculated by solving the diffusive heat 
equation and can be written as
\begin{equation}
	T_s (r) = T_o + \frac{S}{6\kappa \rho C_p} (r_o^2-r^2),
	\label{eq:Ts}
\end{equation}
where $T_o$ is the imposed temperature at the boundary, $r=r_o$. 
The governing equations are solved in dimensionless form, obtained by scaling lengths
with $r_o$, times with $r_o^2/\nu$, and temperature with \mbox{$\nu S r_o^2/(6\rho C_p\kappa^2)$}.
The system of dimensionless equations is:  
\begin{eqnarray}
	&&\pdt{\vel} + \pl \vel \cdot \boldsymbol{\n} \pr \vel + \frac{2}{\Ek}\mathbf{e}_z \times \vel 
	= - \boldsymbol{\n} p + \boldsymbol{\n}^2 \vel + \Ra \Theta r\vect{e}_r ,
	\label{eq:NS1}
	\\
	&& \nabla \cdot \vel = 0,
	\\
	&& \pdt{\Theta} + \vel \cdot \nabla \Theta -\frac{2}{\Pran} r u_r = \frac{1}{\Pran} \nabla^2 \Theta,
	\label{eq:T1}
\end{eqnarray}
where $\vel$ is the velocity field, $p$ the pressure and
$\Theta$ the temperature perturbation relative to the static temperature~(\ref{eq:Ts}). 

The dimensionless numbers are, the Ekman number,
\begin{equation}
	\Ek=\frac{\nu}{\Omega r_o^2}, 
\end{equation}
the Rayleigh number,
\begin{equation}
	\Ra=\frac{\alpha  g_0 S r_o^6}{6 \rho C_p \nu \kappa^2},
\end{equation}
and the Prandtl number,
\begin{equation}
	\Pran=\frac{\nu}{\kappa}.
\end{equation}

At $r=r_o$, the boundary condition for the velocity is no-slip and impenetrable and 
the temperature is fixed, 
\begin{equation}
	\vel = \vect{0}, \quad  \Theta = 0 \textrm{ at } r=r_o.
\end{equation}

Throughout this paper, we use both spherical coordinates $(r,\theta,\phi)$ and 
cylindrical polar coordinates $(s,\phi,z)$.
The mathematical formulation and the numerical method are described in detail in \citet{Guervilly2016},
where the linearised version of the code is benchmarked against theoretical and previous numerical results at the onset of convection. 
The governing equations and the assumptions of the model are briefly described below. 

\subsection{Governing equation for the non-axisymmetric flow}
To model the system of equations~(\ref{eq:NS1})-(\ref{eq:T1}) for small Ekman and Rossby numbers,
we use the quasi-geostrophic approximation to model the evolution of the velocity field 
\citep[\eg][]{Or87,Car94,Gil06}.
The QG approximation reduces the 3D system to a 2D system 
by taking advantage of the small variations of the flow along $z$ compared with variations in $s$ and $\phi$ due to the rapid rotation.
This approximation is only justified in the case of small slope of the boundaries, such as
the \citet{Bus70} annulus. In the case of a sphere, the approximation is therefore
not rigorously justified in any asymptotic limit. 
Consequently, our QG model is intended as a simplified model of convection in a rapidly rotating sphere
that allows us to investigate unexplored regions of the parameter space.
When possible, comparisons with theoretical, experimental and 3D numerical models
show that the QG model correctly reproduces key properties of the full system \citep{Aub03,Mor04,Gil06,Gil07,Pla08}.

The QG model assumes that the fluid dynamics is dominated by the geostrophic balance, \ie the
Coriolis force balances the pressure gradient at leading order. The \revision{leading-order} velocity $\vel^g$
is invariant along $z$ and $\vel^g = (u^g_s,u^g_{\phi},0)$ in cylindrical polar coordinates.
By taking the $z$-component of the curl of the momentum equation~(\ref{eq:NS1}) and averaging it
along $z$, we obtain the equation for the axial vorticity, $\vorz^g = \pl \bnabla \times \vel^g \pr \cdot \vect{e}_z$,
\begin{equation}
	\pdt{\vorz^g} + \pl \vel^g \cdot \bnabla \pr \vorz^g 
	- \pl \frac{2}{\Ek} + \vorz^g \pr \moyz{\frac{\partial u_z}{\partial z}}
	= \nabla_e^2 \vorz^g - \Ra \moyz{\frac{\partial \Theta}{\partial \phi}},
	\label{eq:vorz}
\end{equation}

with 

\begin{equation}
	\nabla^2_e A  \equiv \frac{1}{s} \frac{\partial}{\partial s}\pl s \frac{\partial A}{\partial s} \pr
					+ \frac{1}{s^2} \frac{\partial^2 A}{\partial \phi^2},
\end{equation}
and 
\begin{equation}
 \moyz{A}  \equiv \frac{1}{2H} \int^{+H}_{-H} A dz,
\end{equation}
where $H=\sqrt{1-s^2}$ is the axial distance from the spherical boundary to the equatorial plane.

The velocity $\vel^g$ can be described by a streamfunction $\psi$ that models the 
non-axisymmetric (\ie $\phi$-dependent) components with the addition of an axisymmetric azimuthal flow,
\begin{equation}
	\vel^g = \frac{1}{H} \bnabla \times \pl H \psi \vect{e}_z \pr + \moyp{u_{\phi}^g} \vect{e}_{\phi},
	\label{eq:defPsi}
\end{equation}
where
\begin{equation}
 \moyp{A}  \equiv \frac{1}{2\pi} \int_{0}^{2\pi}A d\phi.
\end{equation}
We choose this formulation for the streamfunction to account for the non-zero divergence of $\vel^g$ in the equatorial plane due
to the return axial flow at the sloping boundaries, 
\begin{equation}
	\bnabla_e \cdot \vel^g = -\beta u_s^g,
\end{equation}
where
\begin{equation}
	\bnabla_e \cdot  \vect{A} \equiv \frac{1}{s} \frac{\partial s A_s}{\partial s} + \frac{1}{s} \frac{\partial A_{\phi}}{\partial \phi},
\end{equation}
and
\begin{equation}
	\beta = \frac{1}{H}\frac{dH}{ds} = - \frac{s}{H^2}.
\end{equation}

The axial velocity $u_z$ is assumed to be linear in $z$. 
The third term on the left-hand side of equation~(\ref{eq:vorz}) requires us to determine 
 $u_z$ at the boundary $z=\pm H$: 
\begin{equation}
	\left. u_z\right|_{\pm H}   = \pm \frac{1}{H} \left. \vel \cdot \mathbf{n}\right|_{\pm H} 
	\pm \beta H u_s^g ,
	\label{eq:uzH}
\end{equation}
where the normal vector at the boundary is $\vect{n} = \vect{e}_r$.
The normal component, $\left. \vel \cdot \mathbf{n}\right|_{\pm H}$, is the Ekman pumping
induced by the viscous boundary layer and is determined by asymptotic methods
for a linear Ekman layer, $ \left. \vel \cdot \mathbf{n}\right|_{z=\pm H}  = \Ek^{1/2} P(s,u_s^g, u_{\phi}^g)$ \citep{Gre68}.
The analytical function $P$ is derived for a spherical boundary in \citet{Sch05}.

The numerical code solves the evolution equation of the non-axisymmetric streamfunction $\psi$.
The no-slip and impenetrable boundary conditions imply that $\psi = \partial_s \psi = 0$ at $s=1$.
\revision{We use the regularity condition $\hat{\psi}^m=\mathcal{O}(s^m)$ at $s=0$, where $\hat{\psi}^m(s,t)$ is the Fourier mode of azimuthal wavenumber $m$
(see \citet{Guervilly2016} for more detail).}

\subsection{Governing equation for the zonal flow}
In our model, the streamfunction $\psi$ only describes the non-axisymmetric motions, so the axisymmetric 
azimuthal flows, or zonal flows, are treated separately.
We take the $\phi$- and $z$-averages of the $\phi$-component of the momentum equation to obtain
\begin{equation}
 \pdt{\moyp{u_{\phi}^g}}
 +   \moyp{u_{s}^g \frac{\partial u_{\phi}^g}{\partial s}} + \moyp{\frac{u_{s}^g u_{\phi}^g}{s}}
 +\frac{2}{\Ek} \left \langle \moyp{u_s} \right \rangle
 = \n^2 \moyp{u_{\phi}^g} - \frac{\moyp{u_{\phi}^g}}{s^2} .
\label{eq:uzonal}
\end{equation}

Note that the geostrophic balance imposes that $\moyp{u_{s}^g}=0$.
The fourth term on the left-hand side of (\ref{eq:uzonal}) involves the $z$-dependent radial velocity, 
which corresponds to the Ekman pumping term. 
Using the incompressibility of the fluid, it can be shown \citep{Aub03} that
\begin{equation}
 \left \langle \moyp{u_s} \right \rangle= 
 	\frac{\Ek^{1/2}}{2H^{3/2}} \moyp{u_{\phi}^g} .
\end{equation}

The no-slip boundary condition at the outer sphere and the symmetry at the centre imply that $\moyp{u_{\phi}^g} = 0$ at  $s=0,1$.

\subsection{Governing equation for the temperature}

The dimensionless equation for the evolution of the temperature perturbation in 3D is
\begin{equation}
 \pdt{\Theta}+ \vel^{3d} \cdot \boldsymbol{\n} \Theta = 
	\frac{1}{\Pran} \pl 2 r u^{3d}_r+\n^2 \Theta \pr.
	\label{eq:T}
\end{equation}
where $\vel^{3d}$ is the velocity in 3D. 
In cylindrical polar coordinates, 
\begin{equation}
	\vel^{3d} = (u_s^g, u_{\phi}^g,\Ek^{1/2} z P + \beta z u_s^g) .
	\label{eq:u3d}
\end{equation}

The temperature is fixed at the outer boundary so $\Theta = 0$ at  $r=1$.
\revision{At the centre of the sphere, the non-spherically symmetric components of $\Theta$ are zero by symmetry 
and the spherically symmetric component of $\partial_r \Theta$ is zero.}
 
\subsection{Numerical method}

In the following, the superscripts $g$ are removed for clarity.
The evolution equations for $\psi$ and $\moyp{u_{\phi}}$ are solved on a 2D grid 
in the equatorial plane. A 
second-order finite difference scheme is implemented in radius with irregular spacing (finer
near the outer boundary). In the azimuthal direction, the variables are expanded in Fourier modes. 
The evolution equation for the temperature is solved on a 3D grid. 
Similarly to the 2D grid, a finite difference scheme is used in radius. 
The temperature is expanded in spherical harmonics $Y_l^m$ in the angular coordinates with $l$
representing the latitudinal degree and $m$ the azimuthal mode. 
\revision{Further detail about the numerical interpolations between the 2D and 3D grids used to compute the buoyancy term and the advection
of the temperature can be found in \citet{Guervilly2016} and \citet{Gue10_thesis}.}

Table~\ref{tab:list} gives the list of the simulations presented in this paper with some output quantities and the numerical resolutions.
To quantify some of the global properties of convection, we often use 
the Reynolds number, which is calculated from the output of the simulations and corresponds to the time-averaged root mean square (r.m.s.) value of the velocity
in dimensionless unit,
 \begin{equation}
	\urms = \frac{1}{\Delta t} \int_{\Delta t}  \pl \frac{3}{4\pi} \int_0^{2\pi} \int_0^{1} (u_s^2 + u_{\phi}^2) 2 H(s) s ds d\phi \pr^{1/2} dt,
	\label{eq:def_Re}
 \end{equation}
where $u_{\phi}$ includes the zonal velocity. 
We measure the convective Reynolds number, $\ucrms$, as in equation~(\ref{eq:def_Re})
but including only the non-axisymmetric velocity.
Similarly, we measure the zonal Reynolds number, $\uzrms$, including only the axisymmetric velocity.
The integration time over which the time averages are calculated is indicated in table~\ref{tab:list} for each simulation.

\begin{table*}
\begin{center}
\begin{tabular}{c c c c c c c c c}
\hline \hline
$\Ek$ & $\Pran$ & $\Ra$ & $\Ra/\Ra_c$ & $\Rey_c$ & $\Rey_0$ & $(N^u_s,M^u_{\textrm{max}})$ & $(N^t_r,M^t_{\textrm{max}},L^t_{\textrm{max}})$ & integration time
\\ \hline 
  $10^{-7}$ & $10^{-1}$ & $6\times10^{9}$ & $1.19$ & $1012$ & $445$ & $(1100,200)$ & $(500,150,150)$ & \revision{$3\times10^5$ $(479)$} \\
  $10^{-7}$ & $10^{-1}$ & $1\times10^{10}$ & $1.99$ & $2070$ & $1110$ & $(1100,200)$ & $(500,150,150)$ & \revision{$3\times10^5$ $(794)$} \\
  $10^{-7}$ & $10^{-1}$ & $2\times10^{10}$ & $3.97$ & $3663$ & $2994$ & $(1200,200)$ & $(500,150,150)$ & \revision{$10^5$ $(414)$} \\
  $10^{-7}$ & $10^{-1}$ & $3\times10^{10}$ & $5.96$ & $5093$ & $5064$ & $(1200,200)$ & $(500,150,150)$ & \revision{$10^4$ $(54)$} \\
  $10^{-7}$ & $10^{-1}$ & $4\times10^{10}$ & $7.95$ & $6243$ & $6840$ & $(1200,200)$ & $(500,150,150)$ &\revision{$10^4$ $(61)$} \\
  $10^{-7}$ & $10^{-1}$ & $5\times10^{10}$ & $9.94$ & $7731$ & $9515$ & $(1500,260)$ & $(600,180,180)$ & \revision{$10^4$ $(73)$} \\
  \hline
  $10^{-7}$ & $10^{-2}$ & $1.9\times10^{9}$ & $1.05$ & $5215$ & $5710$ & $(1000,160)$ & $(400,96,96)$ & \revision{$2\times10^4$ $(90)$} \\
  $10^{-7}$ & $10^{-2}$ & $3\times10^{9}$ & $1.65$ & $9027$ & $11708$ & $(1000,160)$ & $(400,96,96)$ & \revision{$10^4$ $(67)$} \\
  $10^{-7}$ & $10^{-2}$ & $4.8\times10^{9}$ & $2.64$ & $11487$ & $21834$ & $(1200,180)$ & $(400,96,96)$ &\revision{$10^4$ $(82)$} \\
  $10^{-7}$ & $10^{-2}$ & $8.5\times10^{9}$ & $4.68$ & $17548$ & $40009$ & $(1400,200)$ & $(400,96,96)$ & \revision{$10^4$ $(105)$} \\
  \hline
  $10^{-8}$ & $10^{-1}$ & $7.45\times10^{10}$ & $0.96$ & $855$ & $243$ & $(1600,256)$ & $(700,200,200)$ & \revision{$2\times10^6$ $(633)$} \\
  $10^{-8}$ & $10^{-1}$ & $7.8\times10^{10}$ & $1.01$ & $1047$ & $291$ & $(1600,256)$ & $(700,200,200)$ & \revision{$2\times10^6$ $(785)$} \\
  $10^{-8}$ & $10^{-1}$ & $1.5\times10^{11}$ & $1.93$ & $3393$ & $1128$ & $(1800,280)$ & $(700,200,200)$ & \revision{$5\times10^5$ $(400)$} \\
  $10^{-8}$ & $10^{-1}$ & $2\times10^{11}$ & $2.58$ & $4876$ & $1893$ & $(1800,280)$ & $(700,200,200)$ & \revision{$5\times10^5$ $(506)$} \\
  $10^{-8}$ & $10^{-1}$ & $3\times10^{11}$ & $3.87$& $7108$ & $4724$ & $(1800,280)$ & $(700,200,200)$ & \revision{$2\times10^5$ $(286)$} \\
  $10^{-8}$ & $10^{-1}$ & $5\times10^{11}$ & $6.44$ & $9920$ & $9341$ & $(1900,300)$ & $(700,200,200)$ & \revision{$10^5$ $(191)$} \\
  $10^{-8}$ & $10^{-1}$ & $7\times10^{11}$ & $9.02$ &  $12023$ & $13176$ & $(2000,320)$ & $(750,200,200)$ & \revision{$2\times10^6$ $(4453)$} \\
   \hline 
  $10^{-8}$ & $10^{-2}$ & $2\times10^{10}$ & $0.68$ & $7200$ & $4753$ & $(1400,220)$ & $(500,128,128)$ & \revision{$2\times10^5$ $(216)$} \\
  $10^{-8}$ & $10^{-2}$ & $3\times10^{10}$ & $1.01$ & $13498$ & $10598$ & $(1400,220)$ & $(500,128,128)$ & \revision{$10^5$ $(177)$} \\
  $10^{-8}$ & $10^{-2}$ & $5\times10^{10}$ & $1.69$ & $23288$ & $23309$ & $(1500,240)$ & $(500,128,128)$ & \revision{$4\times10^4$ $(108)$} \\
  $10^{-8}$ & $10^{-2}$ & $8\times10^{10}$ & $2.70$ & $33163$ & $43609$ & $(1600,260)$ & $(500,128,128)$ & \revision{$5\times10^4$ $(191)$} \\
\hline \hline
\end{tabular}
\end{center}
\caption{List of input and output parameters for all the simulations presented in the paper.
\revision{$\ucrms$ and $\uzrms$ are the convective and zonal Reynolds numbers respectively.}
The columns labelled $(N^u_s,M^u_{\textrm{max}})$ and $(N^t_r,M^t_{\textrm{max}},L^t_{\textrm{max}})$ give the numerical resolutions on the 2D and 3D grids respectively.
The last column gives the integration time used to compute the time averages \revision{in units of $1/\Omega$ and, in brackets, in units of a convective turnover timescale, $l_c/U_c$, where $U_c$ is the r.m.s. convective velocity (equivalent to $\Rey_c$ in 
our dimensionless units) and $l_c$ is the convective lengthscale computed from equation~(\ref{eq:lc}) and averaged between $0.1\leq s \leq0.8$.}
}
\label{tab:list}
\end{table*}

\section{Structure of the convective and zonal flows and scaling of the velocity}
\label{sec:structure}

\subsection{Radial dependence of the convection}
\label{sec:zonation}

\begin{figure*}
\centering
   \subfigure[]{\label{fig:u_Ra745e8}   
   \includegraphics[clip=true,height=7.2cm]{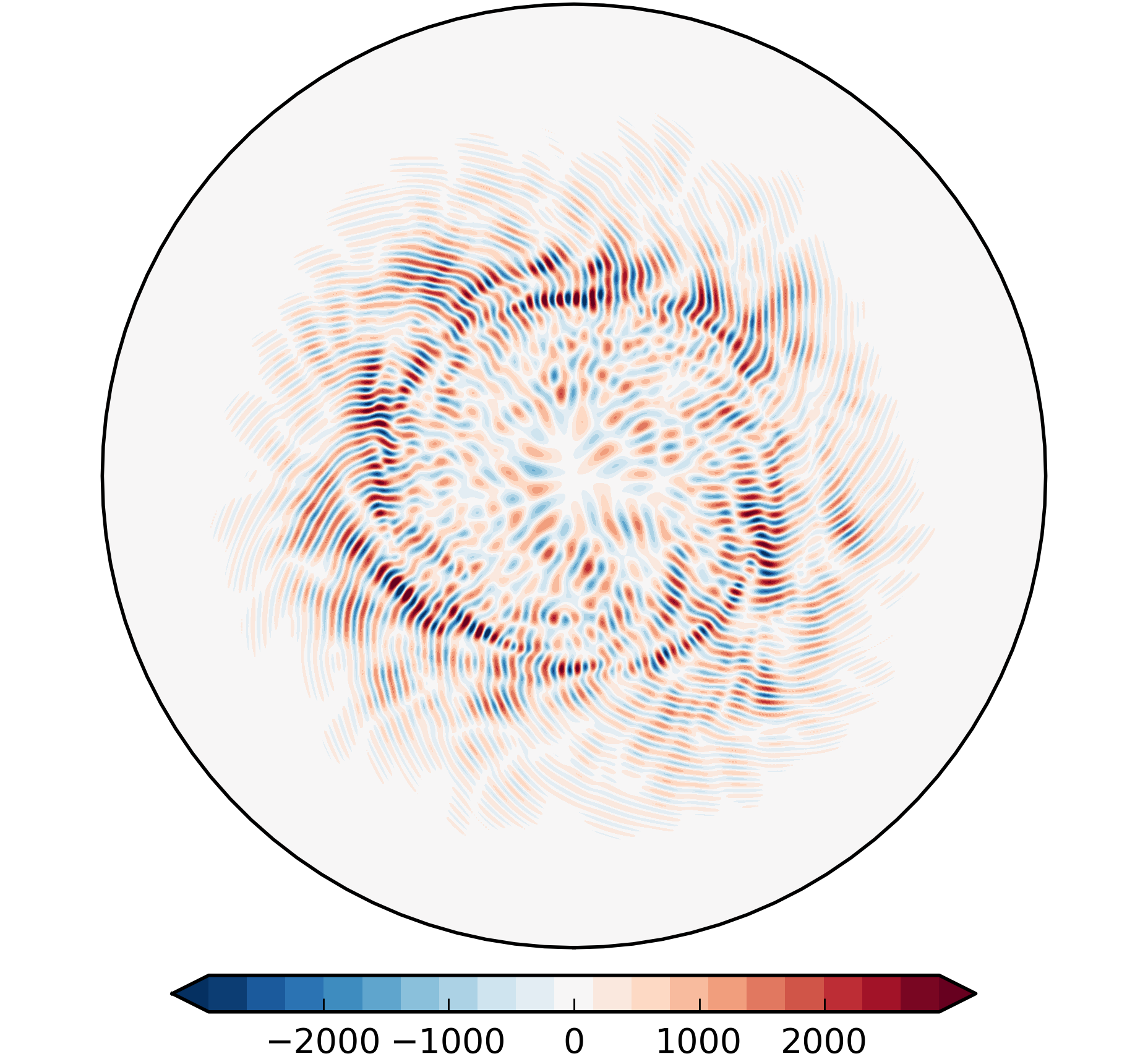}
   \includegraphics[clip=true,height=7.2cm]{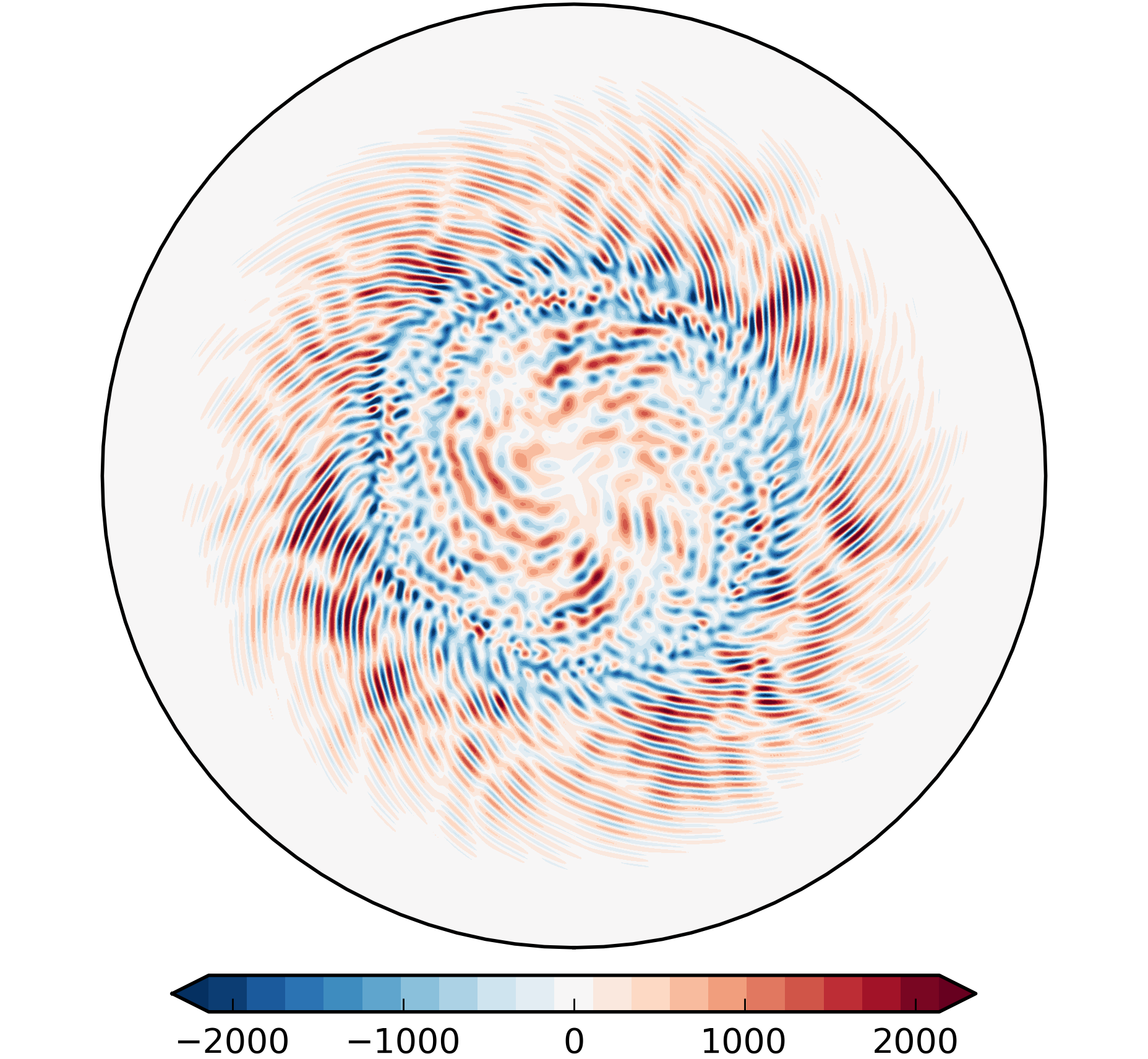}}
   \subfigure[]{\label{fig:u_Ra15e10}   
   \includegraphics[clip=true,height=7.2cm]{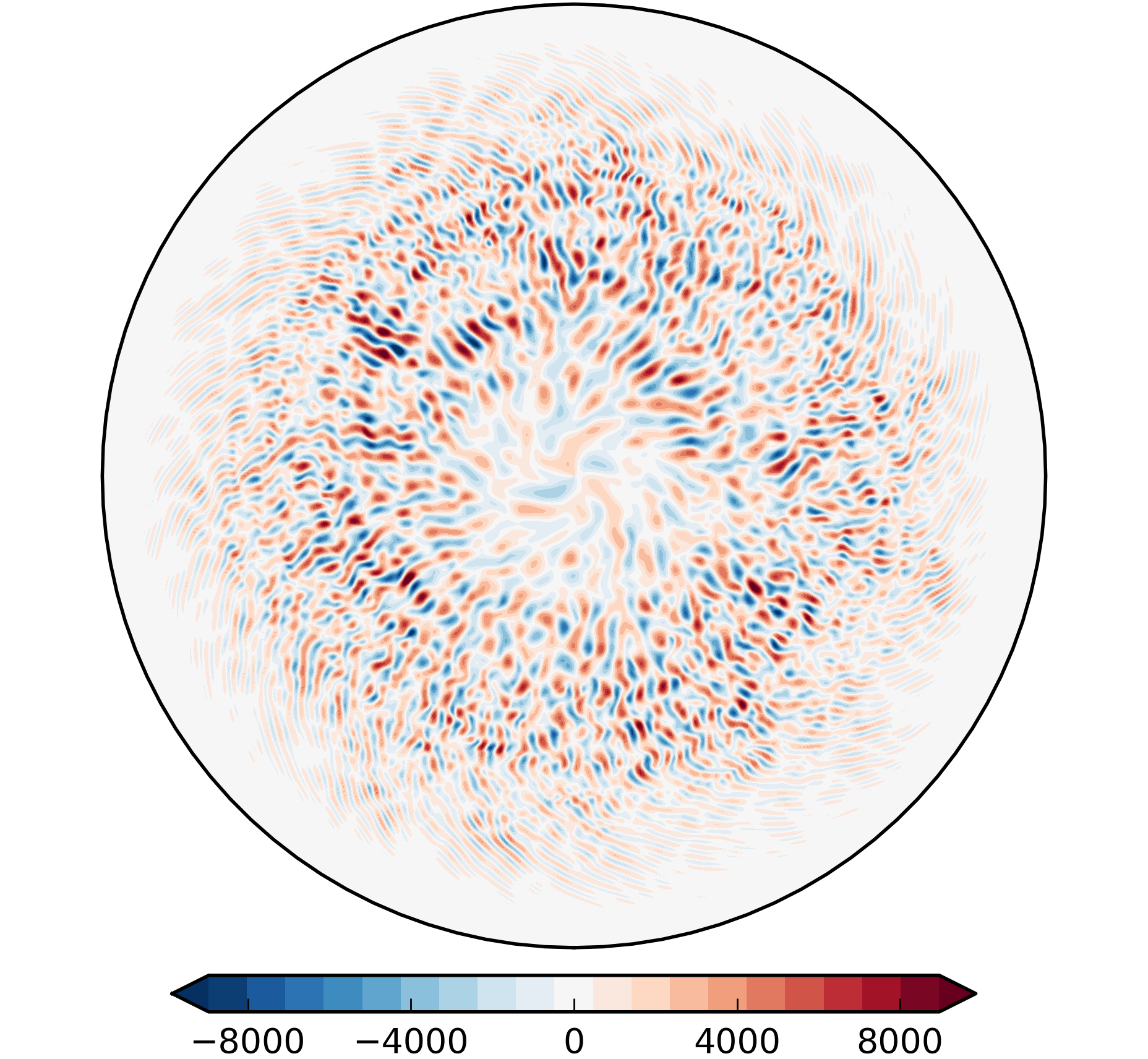}
   \includegraphics[clip=true,height=7.2cm]{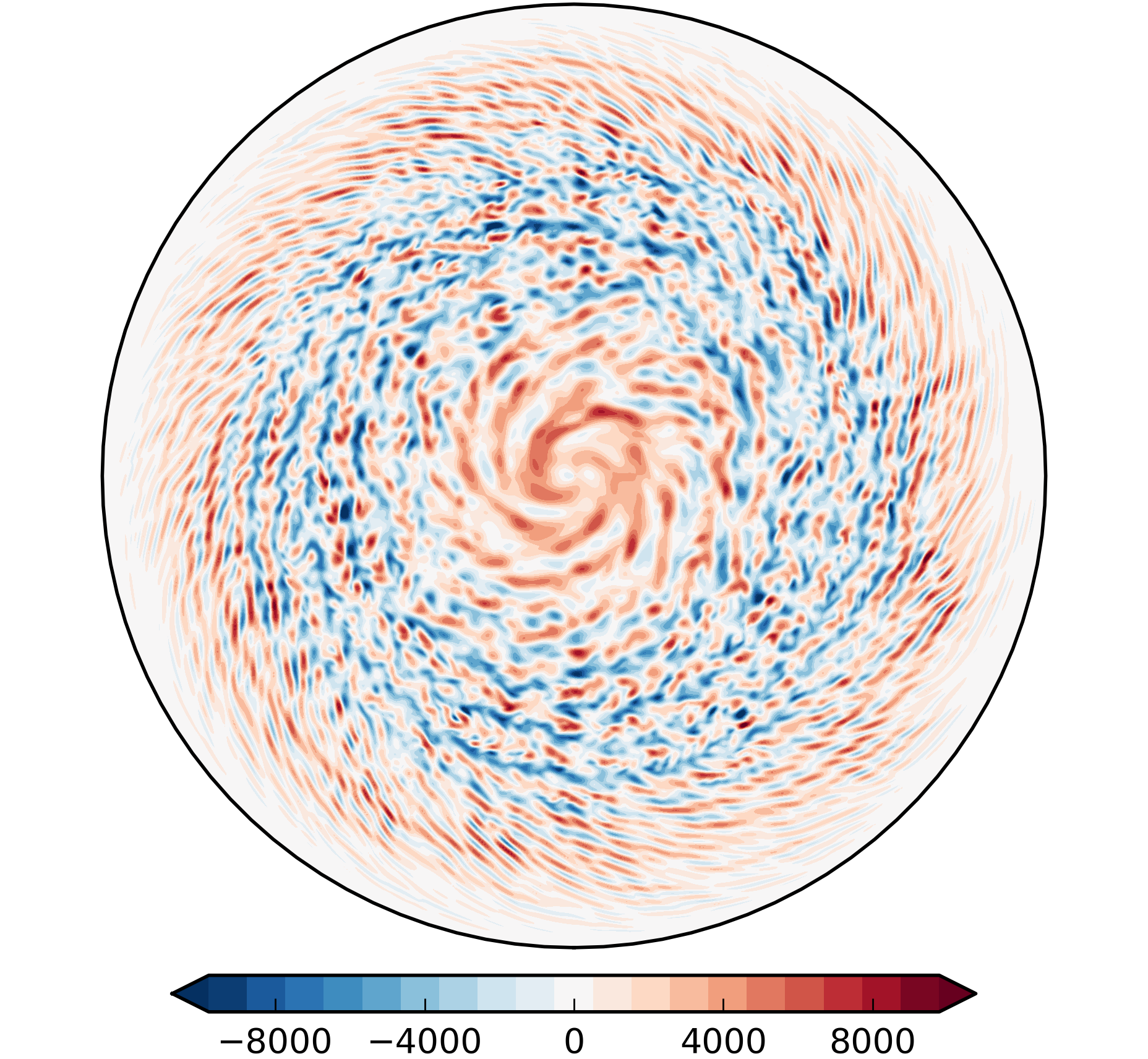}}
   \subfigure[]{\label{fig:u_Ra7e11}   
   \includegraphics[clip=true,height=7.2cm]{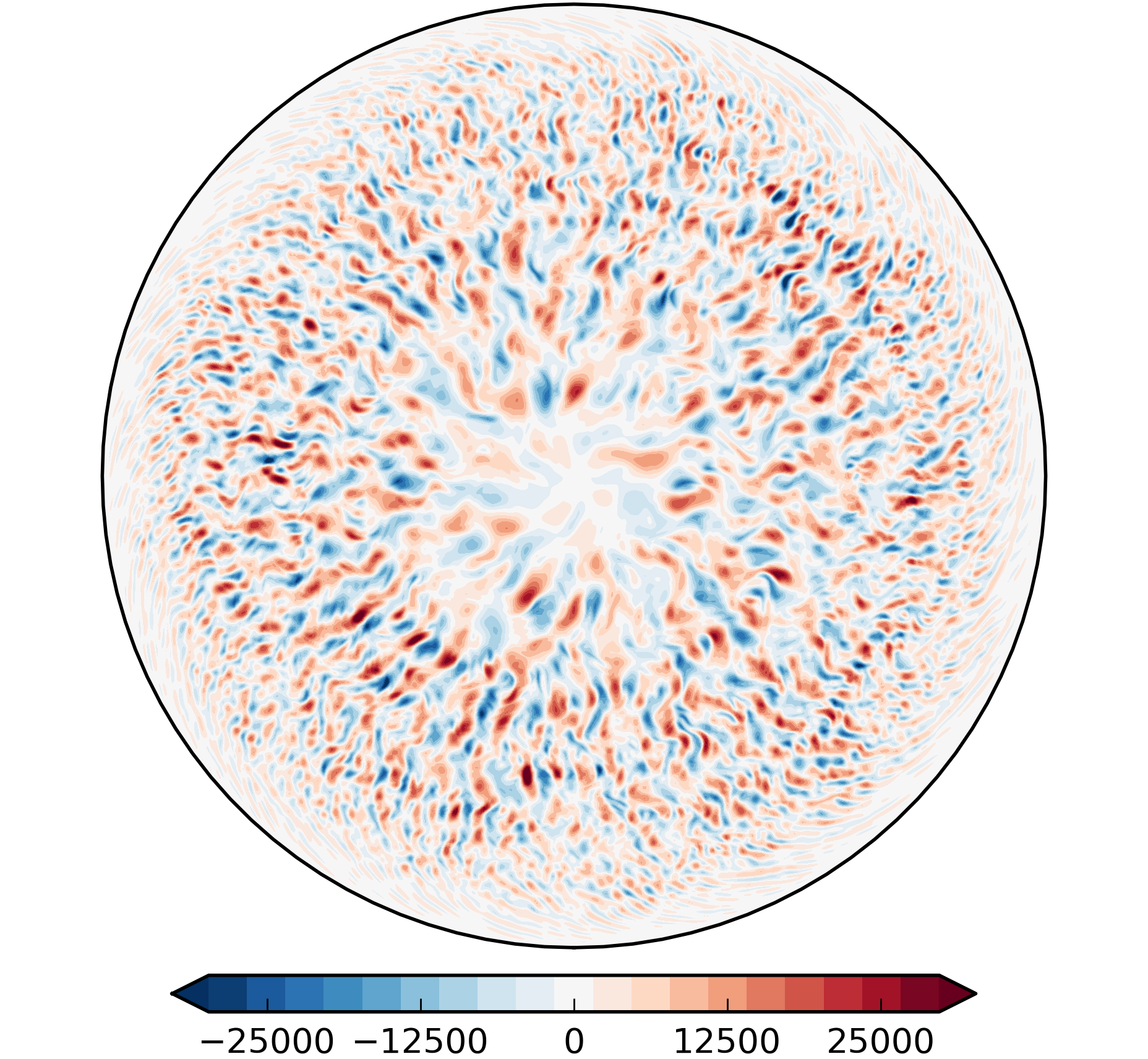}
   \includegraphics[clip=true,height=7.2cm]{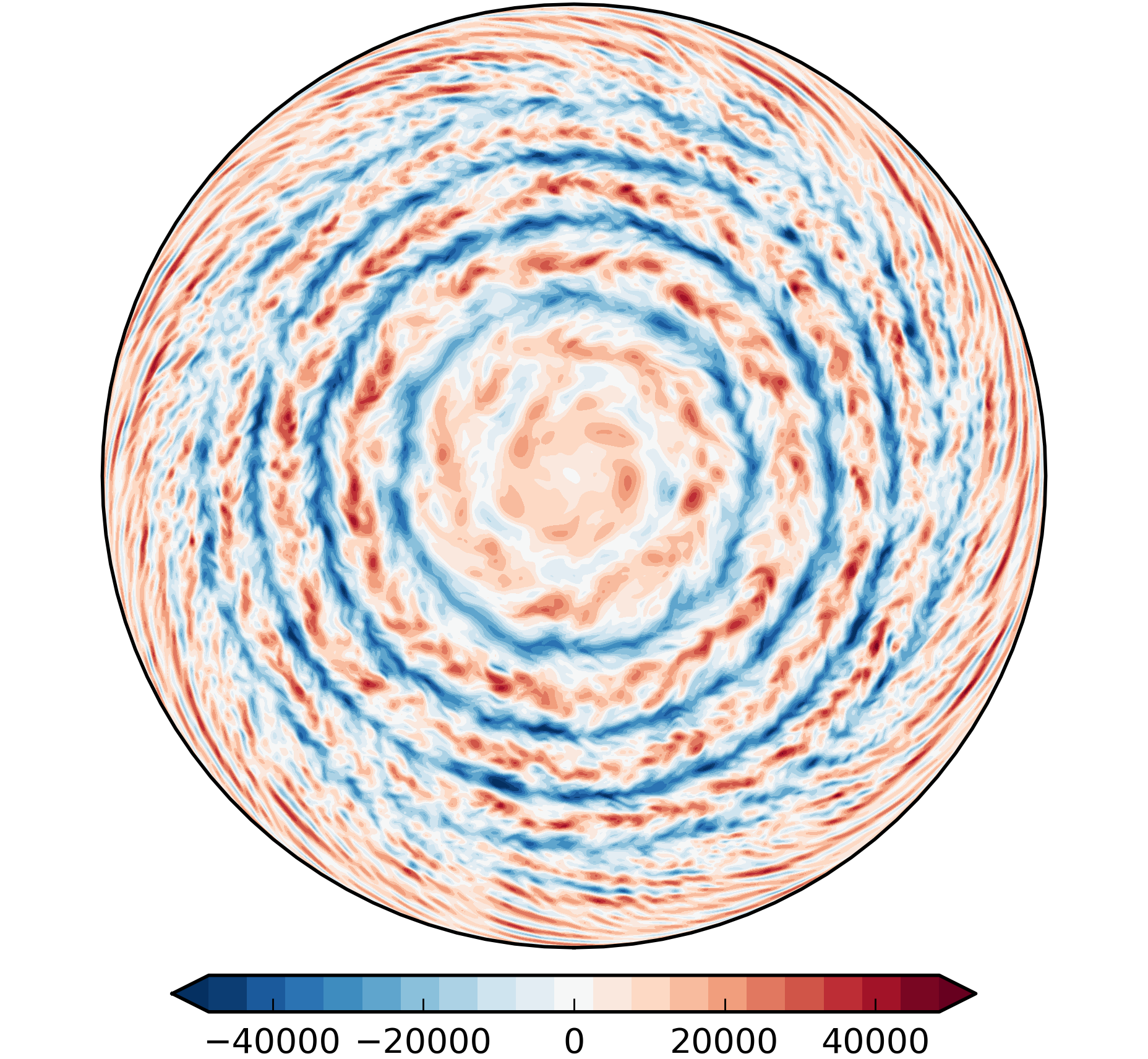}} 
   \caption{Snapshots of the radial velocity (left) and the azimuthal velocity (right)
   in the equatorial plane for $\Ek=10^{-8}$ and $\Pran=10^{-1}$ for (a) $\Ra/\Ra_c=0.96$,
   (b)  $\Ra/\Ra_c=1.93$ and (c) $\Ra/\Ra_c=9.02$.}
   \label{fig:visu_eq}
\end{figure*}

\begin{figure*}
	\centering
	\includegraphics[clip=true,width=0.96\textwidth]{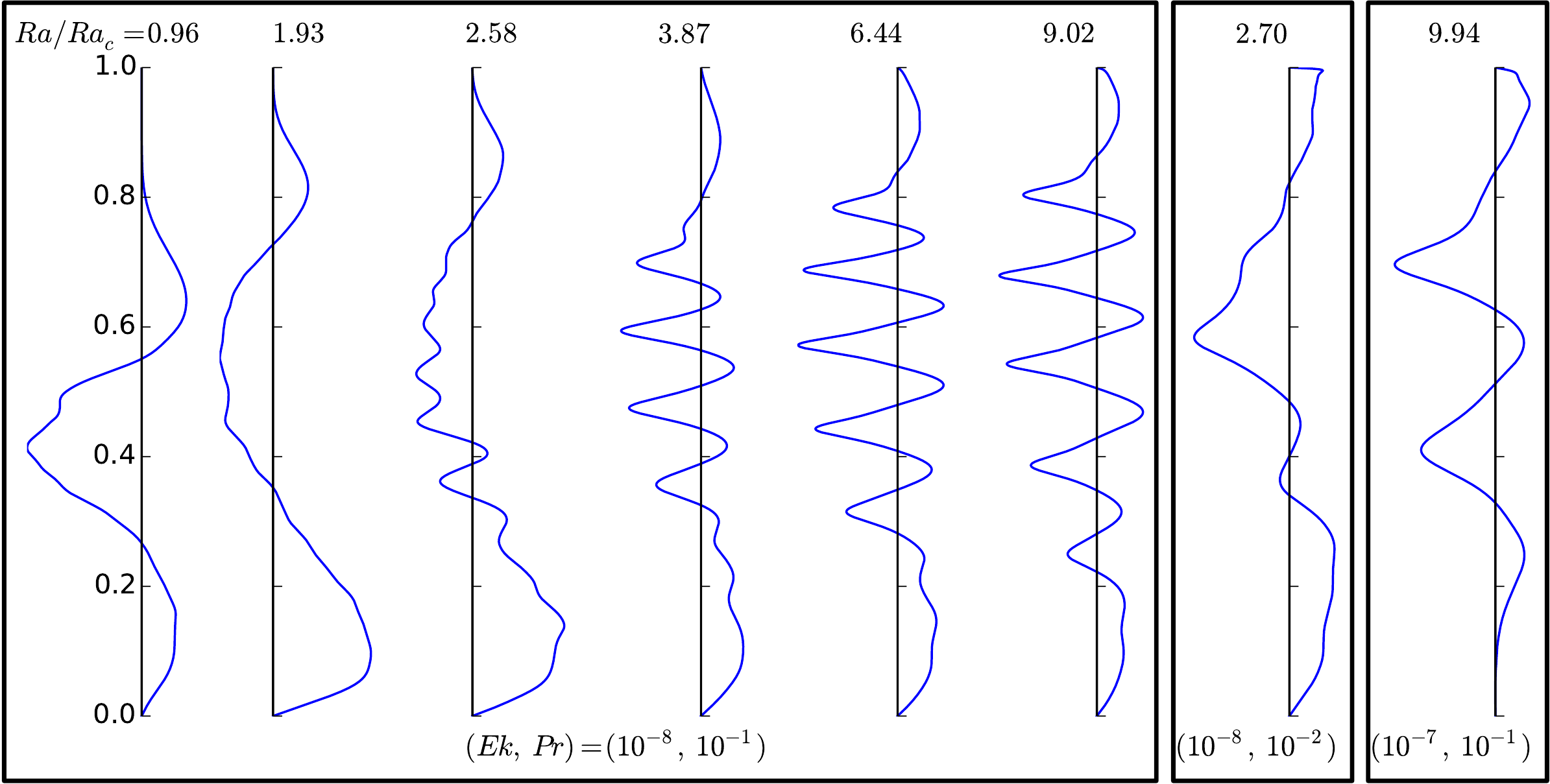}
	\caption{Radial profile of the zonal flow (time average) for different Rayleigh numbers (indicated at the top of each subplot) 
	and Ekman and Prandtl numbers (indicated at the bottom).
	The vertical axis is the radius. The range of the horizontal axis is different for each subplot: 
	the maximum of the zonal flow increases 80-fold between 
	the smallest and largest Rayleigh numbers for $\Ek=10^{-8}$ and $\Pran=10^{-1}$.}
	 \label{fig:profiles}
\end{figure*}

In this section, the Ekman and Prandtl numbers are fixed to $\Ek=10^{-8}$ and $\Pran=10^{-1}$.
For these parameters, the stable solution is located on a strong branch of convection, which is discontinuous 
at the onset of convection \citep{Guervilly2016,Kaplan2017}. All cases presented in this paper are located on the strong branch.
This branch is distinct from the weak branch of convection, which occurs for $\Ek\Pran\gtrsim\mathcal{O}(10^{-8})$
and is continuous at the onset of convection. At the onset of convection, solutions on the weak branch take the form of propagating structures that are tilted 
in the prograde direction. These structures are known as thermal Rossby waves and have been extensively studied in the literature \citep[\eg][]{Bus70,Zha92}. 
Near the onset of convection, the flows on the strong branch are starkly different and are described in detail below.

For $\Ek=10^{-8}$ and $\Pran=10^{-1}$, the nonlinear convection is maintained below the linear onset of convection (quantified by the critical Rayleigh number $\Ra_c$), 
down to a value $\Ra=0.96\Ra_c$ \citep{Guervilly2016}. 
We vary the Rayleigh number from this lower value to approximately $9\Ra_c$. 
Figure~\ref{fig:visu_eq} shows snapshots of the radial and azimuthal velocities in the equatorial plane for $\Ra/\Ra_c=0.96$,  $\Ra/\Ra_c=1.93$ and $\Ra/\Ra_c=9.02$.  
In all three cases, two dynamical regions can be distinguished:
an inner region, where the convection is vigorous with values of the radial velocity up to $3000$ for the lowest $\Ra$ and up to $30000$
for the largest $\Ra$, and an outer region, where the radial flow has smaller amplitude and the flow has finer structures that are tilted in
the prograde direction. This radial dependence of the convection, sometimes referred to as dual convection, was previously described in laboratory experiments \citep{Sum00}
and QG \citep{Aub03} and 3D numerical models \citep{Miy10}.
The limit between the two regions is located around $s=0.5$ for the lowest $\Ra$ and $s=0.8$ for the largest $\Ra$,
so the limit moves outwards when the convection becomes more vigorous.
In the inner region, the convective flows are strongly time dependent, especially for large $\Ra$, and are subject to
frequent nonlinear interactions. 
The contours of the radial velocity tend to be directed radially, contrary to the tilted contours of the outer region. 
For large $\Ra$, the azimuthal lengthscales of the radial flow decreases with increasing radius. This is likely due to
the increase of the slope of the boundary $\beta$ with radius: the vortex stretching term in the axial vorticity equation, 
which depends on $\beta$, impedes the radial motion of wide vortices.  
The azimuthal extent of the convective flows clearly increases with $\Ra$, which indicates the presence of an upscale energy transfer as expected in $\beta$-plane turbulence
 \citep[\eg][]{Davidson2013}.
For all Rayleigh numbers, the radial velocity is weak in the central region because the gravity goes to zero at the centre.
In the outer region, $\beta$ is large and the vortex stretching term is the dominant source of the 
axial vorticity, so this outer region is dominated by the propagation of Rossby waves. 
The nonlinear interactions are weaker in this region.

For all $\Ra$, the azimuthal flow has a visible axisymmetric (\ie zonal) component.
Figure~\ref{fig:profiles} shows the time-averaged profiles of the zonal flow for different $\Ra$.
The zonal flow is prograde in the outermost region for all $\Ra$.
In the outer region dominated by Rossby waves, the Reynolds stresses due to the correlation of the velocity along the tilted contours 
produce a prograde jet in the outer part and a neighbouring inner retrograde jet \citep[\eg][]{Bus82}.  
The behaviour of the zonal flow in the inner convective region is different depending on $\Ra$.
For the lowest $\Ra$ in figure~\ref{fig:visu_eq}, azimuthal flows appear to spiral inward from mid-radius. These flows have an axisymmetric average that is 
positive in the centre and negative near $s=0.4$. The time-averaged profile of the zonal flow for this $\Ra$
has therefore three jets of alternating sign.
For $\Ra/\Ra_c=1.93$, the azimuthal flows consist in a multitude of meandering narrow jets. 
Around mid-radius, the azimuthal average of the narrow meandering jets is not well-defined. Their axisymmetric average
is mostly negative because the retrograde jets have stronger amplitude.
In the centre, the azimuthal flow has a clear prograde direction and is wider than the meandering jets at larger radius. 
For $\Ra/\Ra_c=9.02$, the multiple azimuthal jets have stronger velocity than the radial flow and they do not meander so 
their net axisymmetric average is well-defined. The radial profile of the zonal flow shows persistent multiple jets. 
The central jet remains prograde and is wider than the jets located at larger radius.
Overall, for $\Ra> 3\Ra_c$, the zonal flows develop persistent multiple jets and the innermost and outermost jets always
remain prograde. The region occupied by the in-between jets becomes wider as $\Ra$ increases and the limit between inner convective region 
and outer Rossby wave region moves outwards.
A thin viscous boundary layer can be observed on the profiles of the zonal flow of the largest Rayleigh number
as the boundary condition is no-slip at $s=1$. This thin boundary layer is well resolved in our model as the radial grid is refined near the boundary.

For large $\Ra$, the zonal velocity is larger than the radial velocity. 
In this case, the radial shear exerted by the zonal flow can be faster than the vortex turnover timescale, so
the zonal flow has a dominant role in the dynamics of the convective vortices.
The radial flow has a smaller radial extent at the larger $\Ra$ due to the presence of the multiple zonal jets of strong amplitude. 
The convective structures then change from narrow in $\phi$ and extended in $r$ to fatter in $\phi$ and shortened in $r$  as $\Ra$ increases.

\subsection{Zonal jet width and convective lengthscale}
\label{sec:length}

\begin{figure*}
\centering
    \includegraphics[clip=true,width=14cm]{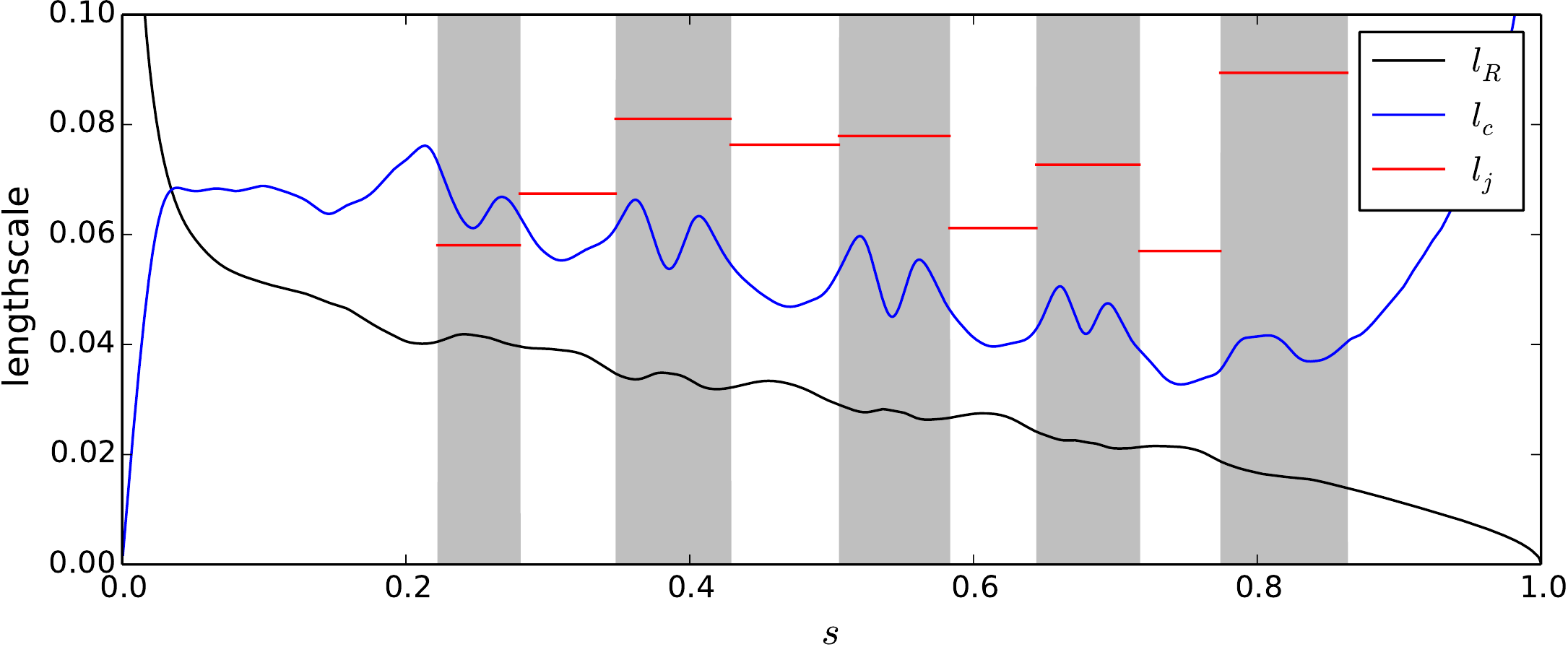}
   \caption{Radial profiles of the Rhines scale $l_R$, the integral scale of the convective flow $l_c$ and the width of the jets $l_j$
   for $\Ek=10^{-8}$, $\Ra/\Ra_c=9.02$ at $\Pran=10^{-1}$.
   The grey bands correspond to the regions where the zonal flow is retrograde.}
   \label{fig:scales}
\end{figure*}

To study the influence of the zonal jets on the convective flow and vice versa, we compare the width of the zonal jet and the convective lengthscale. 
We expect these two lengthscales to be correlated:
on the one hand, the radial shear exerted by the zonal flow on the radial velocity limits the size of the convective flow;
on the other hand, the size of the most energetic convective eddies controls the width of the zonal jet by controlling the mixing length -- this mechanism is discussed in detail in \S\ref{sec:PV}.
To give an estimate of the typical lengthscale of the convection, we compute the integral lengthscale of the non-axisymmetric flow, which is defined as 
\begin{equation}
	l_c(s) = \pi s\pl \frac{\sum\limits_{m=1}^{M^u_{\textrm{max}}} E(m,s)}{\sum\limits_{m=1}^{M^u_{\textrm{max}}} m E(m,s)}\pr,
	\label{eq:lc}
\end{equation}
where $E(m,s)$ is the time-averaged kinetic energy of a mode $m$ of the Fourier decomposition.
The axisymmetric mode ($m=0$) is excluded from this definition.
The convective lengthscale $l_c$ and the width of the jets (denoted $l_j$) are plotted as a function of the radius 
in figure~\ref{fig:scales} for $\Ra/\Ra_c=9.02$.
The regions of retrograde zonal jets are indicated in grey. 
Both the jet width and the convective scale tend to decrease with radius in the inner convective region ($s<0.8$).
$l_c$ takes local minimum values in the core of the jets and maximum values in the flanks of the jets
where the radial shear is largest. $l_c$ corresponds to an azimuthal lengthscale, so this indicates that the shear from the zonal flow elongates 
the convective flow in the azimuthal direction as expected.
The values of $l_c$ and $l_j$ are comparable, although the convective scale is slightly smaller than the jet width.
We have thus verified that a strong correlation exists between the convective scale and the jet width.

In the framework of $\beta$-plane turbulence, a fundamental length scale of the flow is the Rhines scale \citep{Rhi75}, which is given by,
\begin{equation}
	l_{R}(s) = \pi \pl \frac{\Ek U}{2 |\beta|} \pr^{1/2},
	\label{eq:Rhines}
\end{equation}
where $U$ is the 
\revision{r.m.s. velocity of the flow. In the literature, $U$ is interpreted as either an eddy velocity, a jet velocity or a total velocity \citep{Ing82,Dri08}. 
Here we will first consider that $U$ is the r.m.s convective velocity, and later discuss how our results differ when using the total velocity instead.}
Note that $\Ek$ appears in this formula due to our choice of dimensional units.
The Rhines scale may be considered as the scale separating dynamical regimes dominated by either the turbulence (at smaller scales) 
or by Rossby waves (at larger scales). The width of the zonal jets is often thought to be closely related to the Rhines scales, although this notion
has been called into question \citep[\eg][]{Suk07,Scott2012}.
Here we are interested in the predicted dependence of the Rhines scale on $\beta$ and $U$ 
(rather than on the actual value of $l_R$ given by ($\ref{eq:Rhines}$) that arbitrarily includes a factor $\pi/\sqrt{2}$) 
as this dependence can be compared with the lengthscales computed from our data set. 
We first compare the radial dependence of the Rhines scale to $l_c$ and $l_j$.
We plot $l_R$ in figure~\ref{fig:scales}, where we used the r.m.s value of the non-axisymmetric velocity (which varies in radius) to estimate $U$.
In the inner convective region, the convective scale is approximately 2 times larger than the Rhines scale on average 
but the decreasing trends observed for each of the lengthscales is similar: 
between $s=0.3$ and $s=0.7$, $l_j$, $l_c$ and $l_R$ are all approximately divided by two. 
The ratio $l_c/l_R$ is approximately constant in the inner convective region, which shows that
the Rhines scale adequately predicts the radial dependence of the convective scale and the jet width.

As the Rayleigh number (and thus $U$) increases, the Rhines scale predicts that the most energetic convective eddies 
become larger. This is indeed what we observe qualitatively on the snapshots of figure~\ref{fig:visu_eq}.
This increase of the convective scale should be accompanied by an increase of the
jet width. Figure~\ref{fig:profiles} shows that the jets indeed tend to become wider when the 
Rayleigh number increases. However this increase might be
due to the widening of the inner convective region as it pushes the outer region outwards. Larger $\Ra$, currently out of reach of our
computational capabilities,
would be necessary to observe a sizeable increase in the size of the jets.

\begin{figure}
	\centering
	\includegraphics[clip=true,width=8cm]{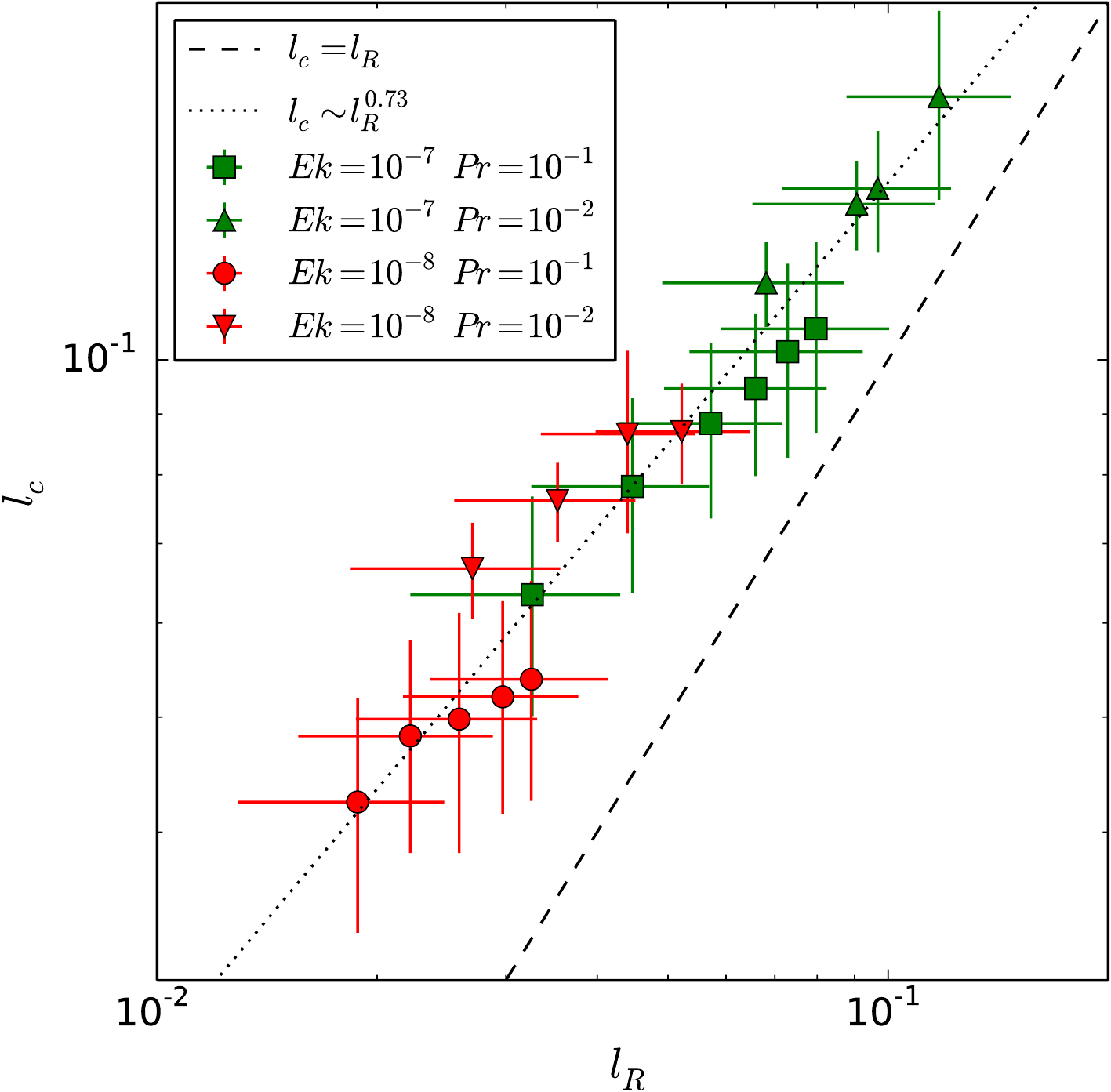}
	\caption{Convective lengthscale versus Rhines scale. Both quantities have been averaged radially in the inner convective region between $0.1\leq s\leq0.8$.
	The horizontal and vertical bars indicate the standard deviation. The dotted line indicates the best fit to all the data points.}
	 \label{fig:scales_all}
\end{figure}
 
We can extend our study on the zonal and convective lengthscales 
and the predictive value of the Rhines scale from the results of simulations performed at different $\Ek$ and $\Pran$.
We cannot presently run simulations at $\Ek<10^{-8}$, so this study is restricted to higher $\Ek$, namely $\Ek=10^{-7}$. 
Our QG-3D model allows us to explore small $\Pran$, so we also use results from calculations at $\Pran=10^{-2}$.  
The two panels on the right of figure~\ref{fig:profiles} shows the zonal velocity for $(\Ek,\Pran)=(10^{-8},10^{-2})$ and
$(\Ek,\Pran)=(10^{-7},10^{-1})$ at the largest $\Ra$ performed (see table~\ref{tab:list}). 
For $(\Ek,\Pran)=(10^{-7},10^{-1})$, the zonal flow has 5 jets of alternating sign. The zonal jets widen when the Ekman number increases if this leads to
a larger r.m.s. convective velocity $\Ek U$ according to the Rhines scale~(\ref{eq:Rhines}).
 
For $(\Ek,\Pran)=(10^{-8},10^{-2})$, the time-averaged zonal flow also has 5 alternating jets. By visual inspection of the snapshots of the velocity, it appears that meandering azimuthal
flows are present 
in the inner region, similarly to the case $\Ra=1.93$ for $(\Ek,\Pran)=(10^{-8},10^{-1})$. 
This suggests that larger Rayleigh numbers would be necessary to get multiple persistent jets. 
However, we were not able to perform calculations at larger $\Ra$ to confirm this. 
Smaller values of $\Pran$ lead to larger values of the convective velocity (see table~\ref{tab:list}), and hence, to wider jets.

To compare more systematically the convective lengthscale with the Rhines scale, figure~\ref{fig:scales_all}
shows the values of $l_c$ versus $l_R$ that have been radially-averaged in the inner convective region (between $0.1\leq s\leq0.8$), 
for increasing values of $\Ra$ and different $\Ek$ and $\Pran$. Both lengthscales vary significantly with radius so we also indicate 
the standard deviation with vertical and horizontal bars.
All the simulations are located on the strong branch of convection, which is discontinuous at the onset of convection, 
and $l_c$ is always larger than the wavelength of the linear instability.
The best fit to all the points is $l_c\sim l_R^{0.73(\pm0.04)}$.
Our simulations therefore indicate that the convective lengthscale increases with the convective flow speed,
but it follows a power law of smaller exponent (namely $l_c\sim U^{0.37}$) than predicted by the Rhines scale (namely $l_c\sim U^{0.5}$).
This result is in agreement with the work of \citet{Gastine2016} using 3D numerical simulations of rotating convection in a spherical shell with $\Pran=1$: 
they find that the convective lengthscale approaches the power law given by the Rhines scale when the Ekman decreases, 
but the exponent remains smaller than a half ($0.45$ for $\Ek=3\times10^{-7}$).
\revision{Here we fitted all the points from different sets of Ekman and Prandtl numbers with one power law because, according to the Rhines scale argument,
the dependence of the convective lengthscale on the parameters can be explained by
a power law dependence on the flow velocity only, irrespective of the values of $\Ek$ and $\Pran$.
However there are some visible variations of the power law exponent for the different data sets, which is further indication that our data do not entirely corroborate the 
Rhines scale argument.}

\revision{We now return to the issue of the interpretation of the velocity scale $U$ in the Rhines scale formula~(\ref{eq:Rhines}).
In our simulations, we find that the convective lengthscale approximately
follows a power law $l_c\sim U^{0.28}$ when $U$ is interpreted as the r.m.s. total velocity. 
The power law exponent is therefore further away from the Rhines scale prediction when using the total velocity rather than the convective velocity.}

\subsection{Scaling of the zonal flow velocity}
\label{sec:scaling}

\begin{figure*}
	\centering
	\includegraphics[clip=true,width=7.8cm]{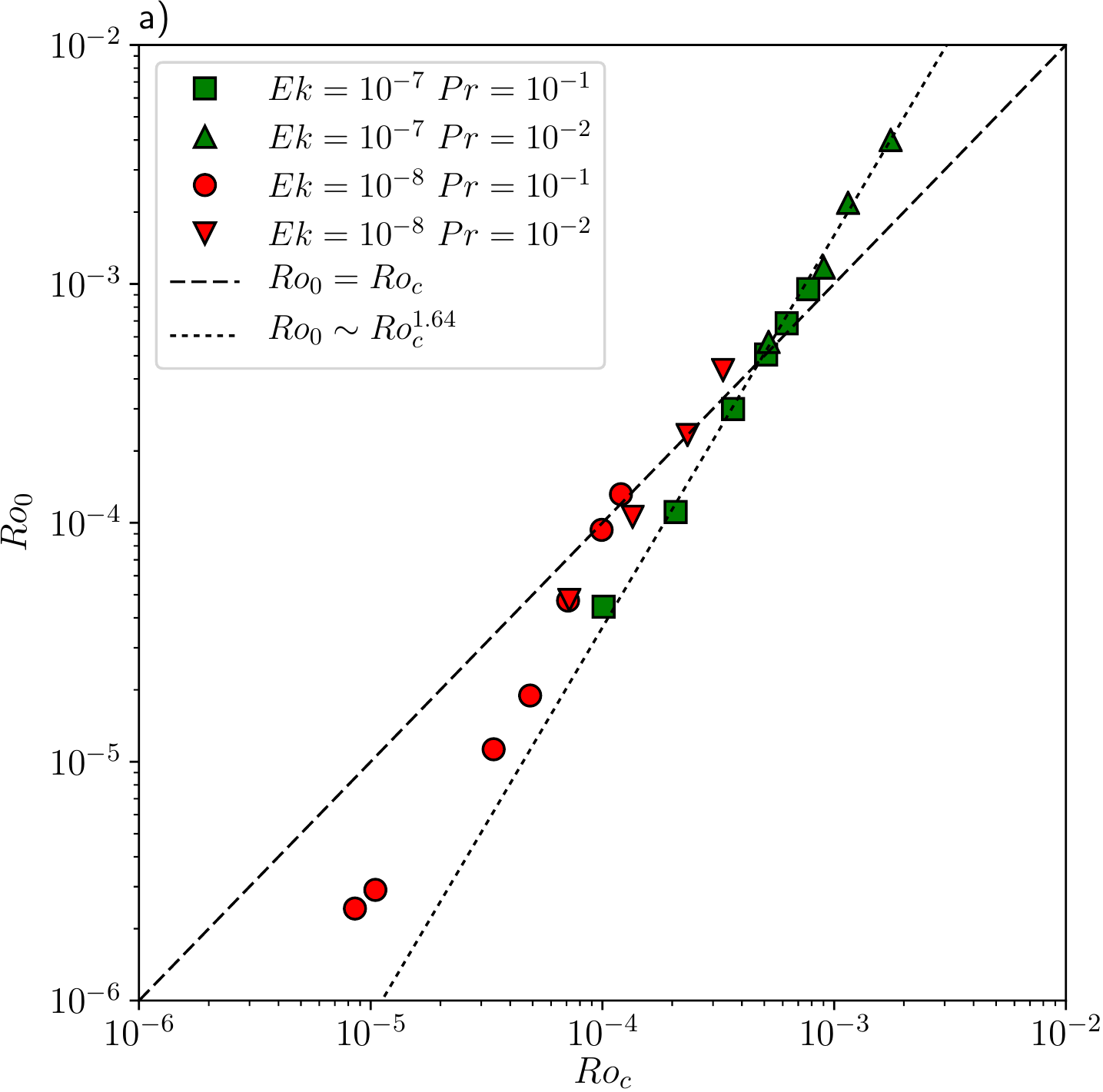}
	\includegraphics[clip=true,width=7.8cm]{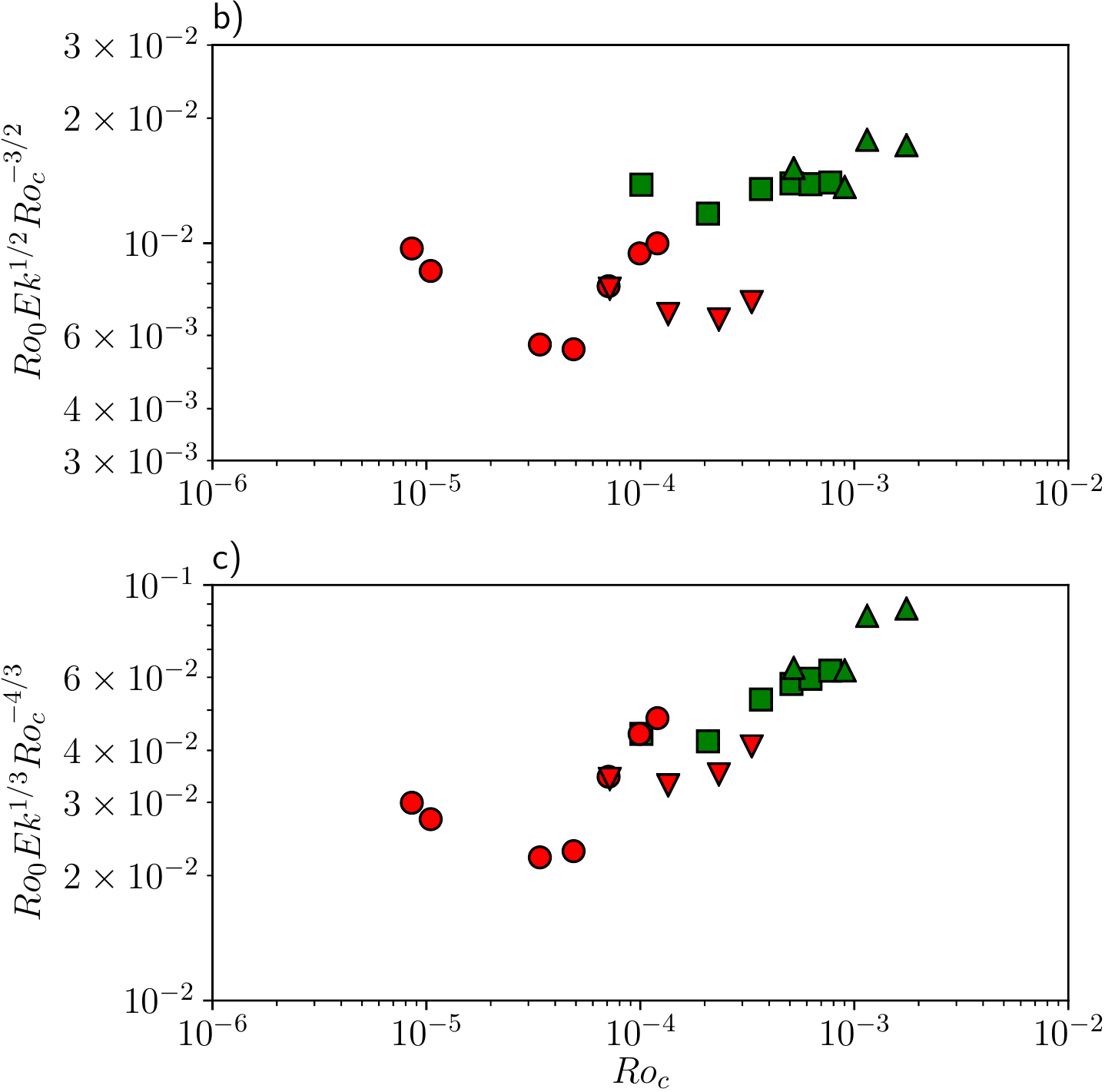}
	\caption{\label{fig:UcUzon_Ro}
	(a) Zonal Rossby number $\Roz$ as a function of the convective Rossby number $\Roc$.
	\revision{The dotted line shows the power law whose exponent best fits all the data points.}
	\revision{(b)-(c) $\Roz$ compensated by the power laws of (b) equation~(\ref{eq:Roc32}) and (c) equation~(\ref{eq:Roc43})
	as a function of $\Roc$.}
	}
\end{figure*}

To complete this section, we now discuss how the amplitude of the zonal velocity scales with the convective velocity.
Figure~\ref{fig:UcUzon_Ro}a shows the evolution of the zonal Rossby number, $\Roz=\uzrms\Ek$, as a function of the convective
Rossby number, $\Roc=\ucrms\Ek$, for $\Ek\in[10^{-8},10^{-7}]$ and $\Pran=[10^{-2},10^{-1}]$. 
In this section we use the Rossby numbers to clarify the dependence of the velocity on the viscosity.
For $\Ek\Pran\leq10^{-9}$, the nonlinear onset of convection is subcritical \citep{Guervilly2016}. 
In this case, the zonal and convective Reynolds numbers are discontinuous and $\ucrms\gtrsim10^3$.
\revision{For each fixed value of $\Ek$, the data points approximately fall on a straight line, indicating a power law dependence on $\Ro_c$.}
The dashed line represents $\Roz=\Roc$. The data points cross this line for moderate values of the Rayleigh numbers
that depend on the Prandtl number: at $\Ra\approx1.5\Ra_c$ for $\Pran=10^{-2}$ and $\Ra\approx6\Ra_c$ for 
$\Pran=10^{-1}$.
The zonal flows have therefore large amplitude for relatively modest values of the Rayleigh numbers when 
$\Pran<1$, despite the presence of the Ekman boundary friction in our model. They reach an amplitude 
comparable to the amplitude of the convective flows for smaller $\Ra$ as $\Pran$ is decreased.
In this sense, lower $\Pran$ is favourable for the zonal flows.

A scaling law for the zonal flow amplitude can be obtained by considering that the dominant force balance in the zonal 
velocity equation (\ref{eq:uzonal}) is established between the nonlinear interactions of the convective velocities
and the friction in the Ekman layer. This is consistent with measurements of the contributions to the zonal energy budget in our simulations.
For instance, in the case $\Ek=10^{-8}$, $\Pran=10^{-1}$ and $\Ra/\Ra_c=9.02$, we measured that the boundary friction accounts for $90\%$ of the total
viscous dissipation of the axisymmetric flow. In terms of scaling arguments, this balance implies
\begin{equation}
	 \Roz \sim \Ek^{-1/2} \frac{\Roc^2}{l_c} .
	 \label{eq:RozRoc}
\end{equation} 
Using the scaling $l_c\sim\revision{\Roc}^{0.37}$ deduced from our data in \S\ref{sec:length}, we obtain the scaling $\Roz\sim \Roc^{1.63}$. 
The best fit to all the data points in figure~\ref{fig:UcUzon_Ro}a is $\Roz\sim \Roc^{1.64(\pm0.04)}$. This agreement is not surprising because all the global quantities 
are calculated from the same data set, but it shows that our measurements of the global values of the convective lengthscales and the 
velocities are consistent \revision{and that equation (\ref{eq:RozRoc}) is suitable to scale the zonal flow velocity}. 

It is interesting to compare the observed scaling of the zonal velocity with predictive scaling laws based on physical arguments
that have been derived in the literature \citep[\eg][]{Aub01}.
To do so, we need to examine possible scaling arguments for $l_c$.
Considering first that $l_c$ scales with the Rhines scale ($\ref{eq:Rhines}$), $l_c\sim \Roc^{1/2}$, gives
\begin{equation}
	\Roz \sim \Ek^{-1/2} \Roc^{3/2}.
	\label{eq:Roc32}
\end{equation}
The exponent $3/2$ is in good agreement with the results of the 3D numerical simulations of \citet{Kaplan2017} for 
$\Ek\in[10^{-7},10^{-6}]$ and $\Pran\in[3\times10^{-3},10^{-1}]$.

\citet{Gil07} proposed an alternative scaling for $l_c$ obtained by using the zonal velocity as a typical flow velocity in the Rhines scale.
This gives $l_c\sim\Roz^{1/2}$, and so, 
 \begin{equation}
	\Roz \sim \Ek^{-1/3} \Roc^{4/3}.
	\label{eq:Roc43}
\end{equation}
\citet{Gil07} found that this scaling law provides a good fit for their numerical data obtained with a QG model at $\Ek=\mathcal{O}(10^{-6})$ and $\Pran=0.025$.

Finally, we can consider $l_c\sim\Ek^{1/3}$, which corresponds to the scaling of the azimuthal lengthscale of the linear convective instability \citep{Jon00}.
This estimate is obviously not satisfactory because the convective lengthscale must increase with the Rayleigh number
as shown in figure~\ref{fig:visu_eq} and figure~\ref{fig:scales_all}.
For this estimate, we deduce the scaling law
\begin{equation}
	 \Roz \sim \Ek^{-5/6} \Roc^2.
	 \label{eq:Roc2}
\end{equation} 

The power law (\ref{eq:Roc32}) provides the closest exponent to our data best fit and is based on plausible physical arguments.
By contrast, the power law (\ref{eq:Roc2}) is too steep and the dependence of the prefactor on $\Ek$ is much weaker in the data.
The power law (\ref{eq:Roc43}) requires a much stronger dependence of the convective lengthscale on $\Roc$ (namely \revision{$l_c\sim\Roc^{2/3}$}) than observed.
\revision{Figures~\ref{fig:UcUzon_Ro}b and c show the dependence of $\Roz$ compensated by the power laws (\ref{eq:Roc32}) and (\ref{eq:Roc43}), respectively, 
on $\Roc$. The data points compensated by the power law (\ref{eq:Roc32}) align on a plateau, confirming that the exponent of this power law provides
a reasonable agreement with our data.}

The prefactor of the power law (\ref{eq:Roc32}) predicts that a decrease of $\Ek$ by a decade should lead to an increase of $\Roz$
by approximately a factor 3. \revision{The dependence on the Ekman number cannot be estimated accurately from our data points as
they only sample one decade of $\Ek$. 
Figure~\ref{fig:UcUzon_Ro}b tentatively suggests that the power law (\ref{eq:Roc32}) does not entirely explain the dependence on the Ekman number: 
in our simulations, $\Roz$ only increases by approximately a factor 2 when $\Ek$ decreases from $10^{-7}$ to $10^{-8}$.}

\section{Drift and stability of the zonal jets}
\label{sec:stability}

\begin{figure*}
	\centering
	\subfigure[]{
	\includegraphics[clip=true,width=13cm]{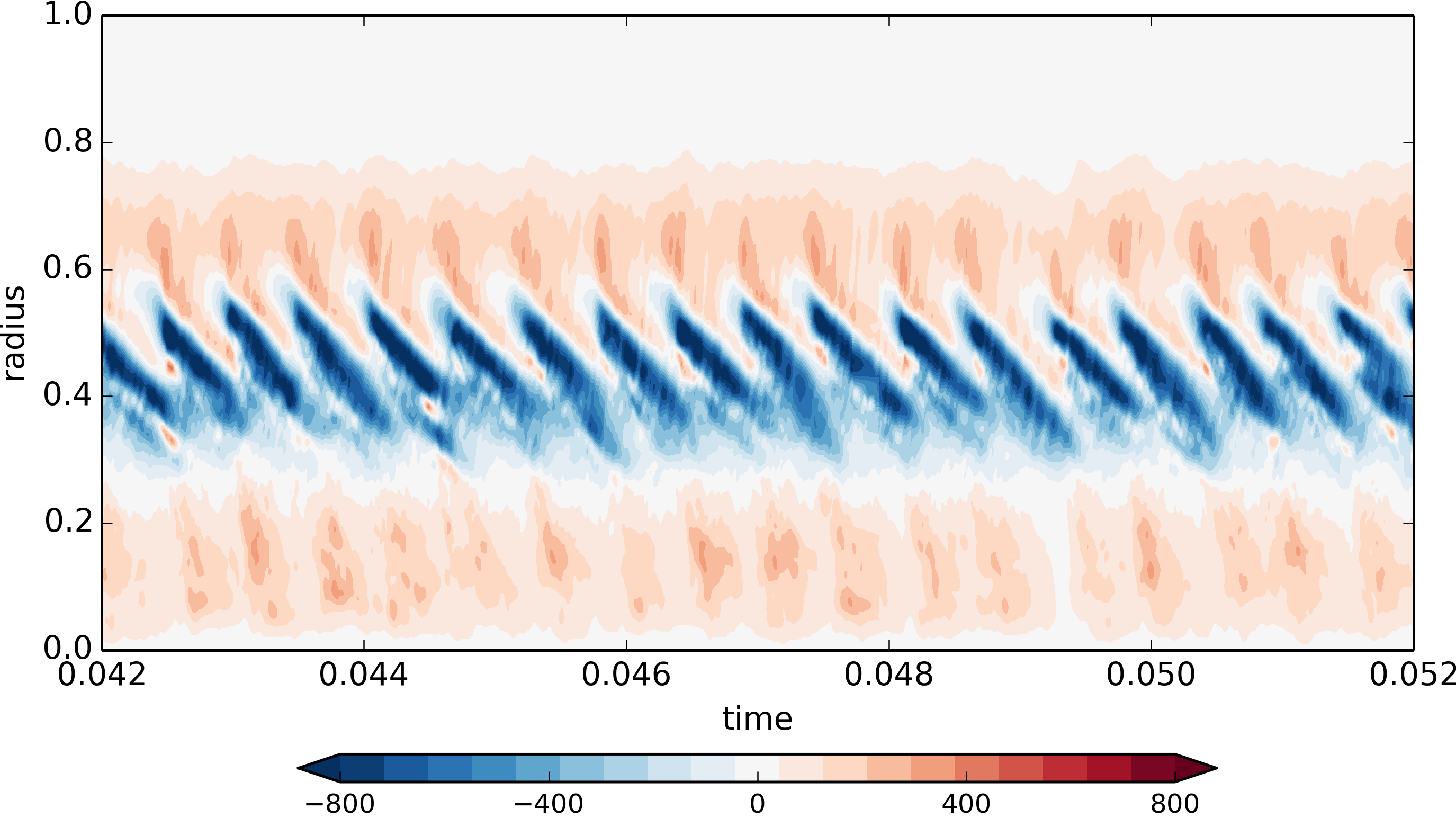}}
	\subfigure[]{ \label{fig:diag_U0b}
	\includegraphics[clip=true,width=13cm]{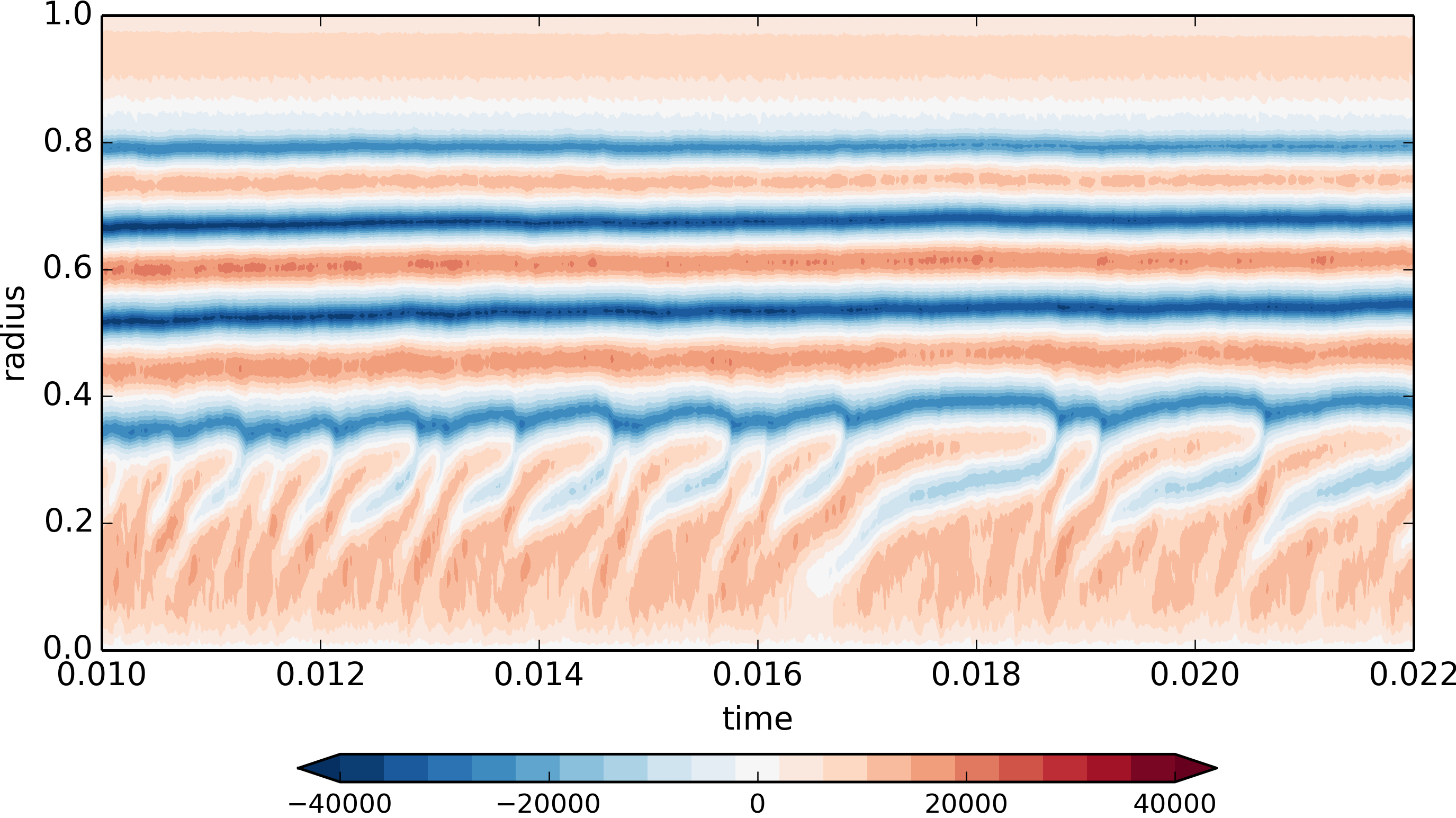}}
	\caption{Space-time diagram of the zonal velocity for (a) $\Ra/\Ra_c = 0.96$ and (b) $\Ra/\Ra_c = 9.02$ ($\Ek=10^{-8}$ and $\Pran=10^{-1}$).
	\revision{The time interval corresponds to (a) $10^6$ rotation periods and $317$ convective turnover timescales (as defined in table~\ref{tab:list})
	and (b) $1.2\times10^6$ rotation periods and $2672$ convective turnover timescales.}}
	 \label{fig:diag_U0}
\end{figure*}

In the rest of the paper, we only consider simulations run at $\Ek=10^{-8}$ and $\Pran=10^{-1}$, where we obtain the largest number of zonal jets.
The zonal flows have a strong influence on the convective flows, and hence on the heat transport as we will discuss in \S\ref{sec:barrier}, so it is important
to discuss the persistence of the zonal jets. In this section, we examine the stability of the jets. 

Using a QG model of thermal convection, \citet{Rotvig2007} shows that zonal jets drift inwards provided that the
slope $\beta$ has a significant dependence on radius,
\revision{while QG models with constant $\beta$ produce multiple jets that do not drift \citep[\eg][]{Jon03}.}
\citeauthor{Rotvig2007} showed that the drift also occurs in a 3D spherical model,
so this effect is not restricted to QG models. 
The drift rate is found to increase with $\beta$ and with the Rayleigh number.
The drift of zonal flows is also observed in the rotating turntable experiment of \citet{Smith2014}. In the experiment, $\beta$ is positive and 
the drift is observed outwards. 
These studies thus indicate that the direction of the zonal jet drift is related to the sign of $\beta$. 

Figure~\ref{fig:diag_U0} shows the space-time diagram of the zonal velocity for $\Ra/\Ra_c=0.96$ and
$\Ra/\Ra_c = 9.02$. 
For $\Ra=0.96$, the middle retrograde zonal jet drifts inwards periodically. 
In our system $\beta<0$ so this inwards migration of the zonal flows is in agreement with the work of \citet{Rotvig2007} and \citet{Smith2014}.

For $\Ra/\Ra_c=9.02$, the picture is completely different. For $s>0.4$ the zonal jets do not drift and the standard deviation of their amplitude is of the order of $2000$ 
(compared with a mean amplitude of approximately $30000$) over the course of the simulation. 
However at radius $s<0.4$, the central prograde jet and its retrograde neighbour drift outwards.
The retrograde jet initially forms around $s=0.2$ and eventually merges with the retrograde jet located at $s\approx0.35$, closing off the prograde jet in the process. 
This sequence is not quite periodic and takes between 5 to 20 zonal turnover timescale
(based on the time- and volume-averaged zonal velocity).
The central region of the equatorial plane has the smallest values of $\beta$ and $d\beta/ds$, so the direction and the location of the drift indicates that this mechanism is
different from the inward drift observed for smaller $\Ra$ and large $\beta$. 

\begin{figure*}
	\centering
	\includegraphics[clip=true,width=13cm]{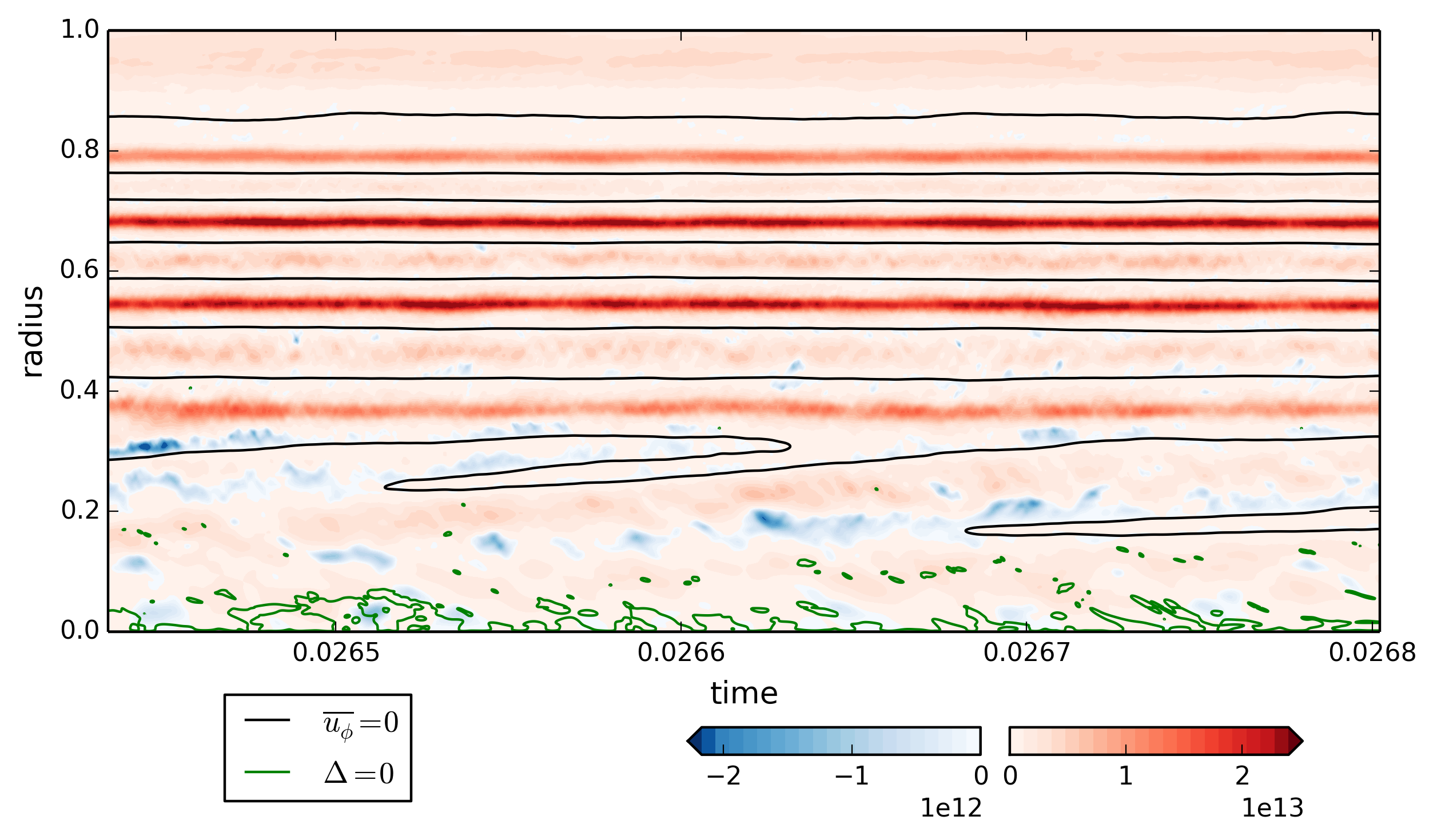}
	\caption{Space-time diagram of $R$ (colour) for $\Ra/\Ra_c = 9.02$ ($\Ek=10^{-8}$, $\Pran=10^{-1}$). 
	The isocontours $\zw=0$ and $\Delta=0$ have been represented in black and green respectively.
	\revision{The time interval corresponds to $3.5\times10^4$ rotation periods and $78$ convective turnover timescales (as defined in table~\ref{tab:list}).}}
	 \label{fig:stab_U0}
\end{figure*}

The drift might be related to instabilities of the prograde jets near the centre, \revision{where $\beta$ goes to zero}. 
The Rayleigh-Kuo criterion states that a necessary condition for the barotropic instability of a shear flow in a inviscid Boussinesq
fluid is that the quantity \mbox{$\Delta=2\beta\Ek^{-1}-d\moyp{\vorz}/ds$}, where \mbox{$\moyp{\vorz}=d\zw/ds+\zw/s$}, 
changes sign at some radius \citep{Kuo49}.
This indicates that large slopes have a stabilising effect on the zonal flow, so it is plausible that prograde zonal flows near the centre
are unstable.
The criterion is valid for an inviscid fluid, so it only provides an indication of the zonal flow stability for small Ekman numbers. 
Nevertheless, \cite{Gue12a} showed that the threshold of the instability obtained with numerical simulations is in good quantitative agreement
with the Rayleigh-Kuo criterion for $\Ek<10^{-7}$.
As well as calculating the quantity $\Delta$ in the time series of $\Ra/\Ra_c=9.02$, we also compute the product of $\zw$ with the zonal average of the 
eddy momentum flux convergence
given in the zonal velocity equation (\ref{eq:uzonal}),
\begin{equation}
 R = - \pl \moyp{u'_{s} \frac{\partial u'_{\phi}}{\partial s}} + \moyp{\frac{u'_{s} u'_{\phi}}{s}} \pr \zw,
\end{equation}
where the prime denotes the non-axisymmetric velocity component.
$R$ is the only source of energy of the zonal flow, so we expect this term to be positive when the eddies feed energy into the zonal velocity.
Figure~\ref{fig:stab_U0} shows the space-time diagram of $R$ (colour), where the isocontours of $\zw=0$ (black line) and $\Delta=0$ (green line) have been superposed.
In the central region, $R$ is negative in the inward flank of the drifting prograde jet, meaning that the zonal flows lose energy to the non-axisymmetric flows there.
The instability criteria (\ie $\Delta>0$) is violated near the centre where $\beta$ is small and in a few places in the drifting prograde jet. 
This indicates that this prograde jet might be \revision{marginally stable}, 
leading to a transfer of energy from the zonal flow to non-axisymmetric flows.
As the zonal flow loses energy in the inward flank of the prograde jet, we expect the fluid parcels located there to move outwards to conserve their angular momentum. 
This mechanism could explain why the prograde jet moves outwards, and by doing so, pushes the neighbouring retrograde jet outwards.

\section{Dynamical difference between retrograde and prograde jets}
\label{sec:PV}

\begin{figure*}
	\centering
	\includegraphics[clip=true,width=14cm]{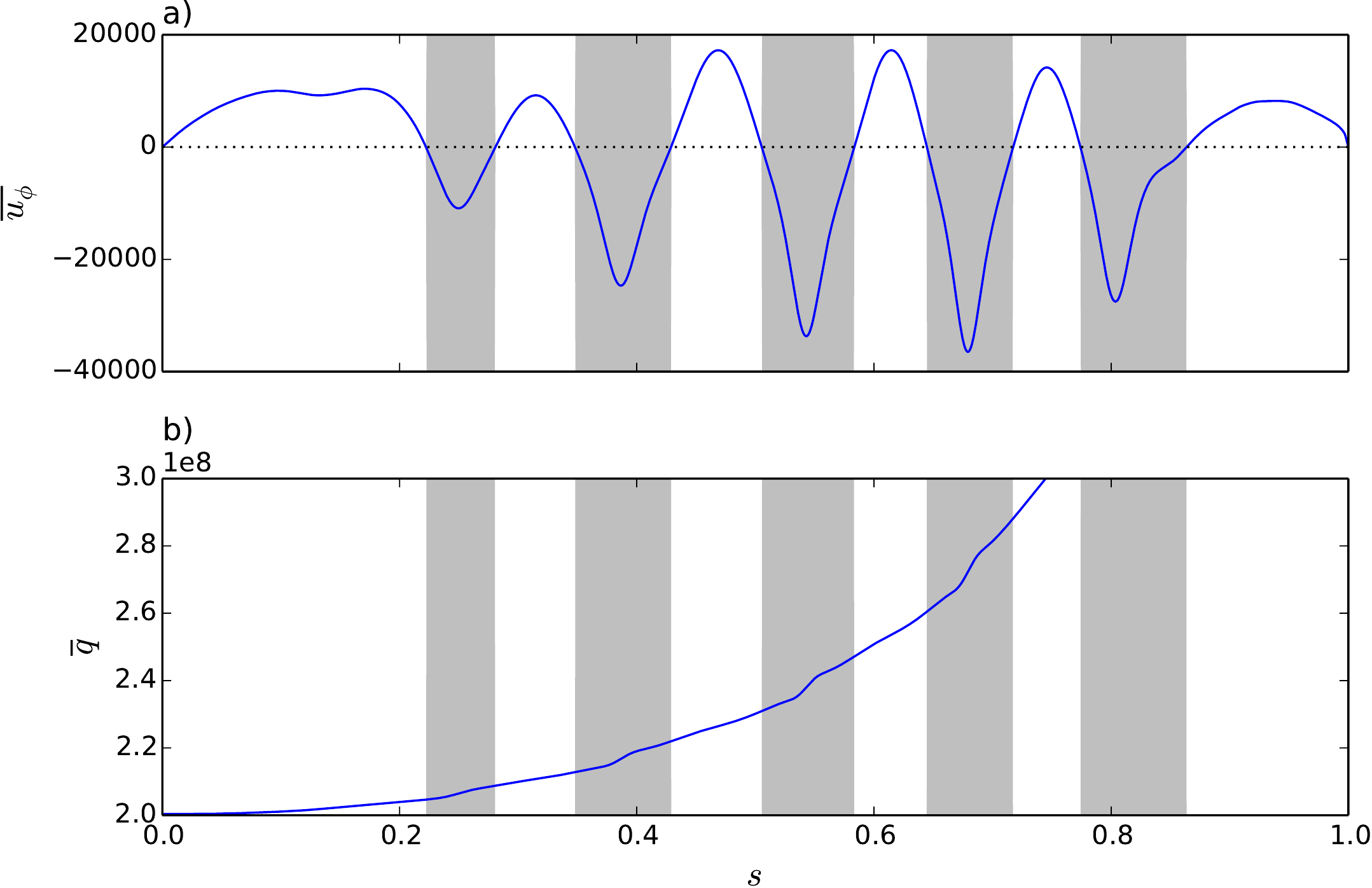}
	\caption{Radial profile of (a) the zonal velocity and (b) the axisymmetric potential vorticity (time averages) for $\Ra/\Ra_c=9.02$ ($\Ek=10^{-8}$, $\Pran=10^{-1}$).
	The grey bands correspond to the regions where the zonal flow is retrograde.}
	 \label{fig:pv_staircase}
\end{figure*}

\begin{figure*}
\centering
    \includegraphics[clip=true,width=14cm]{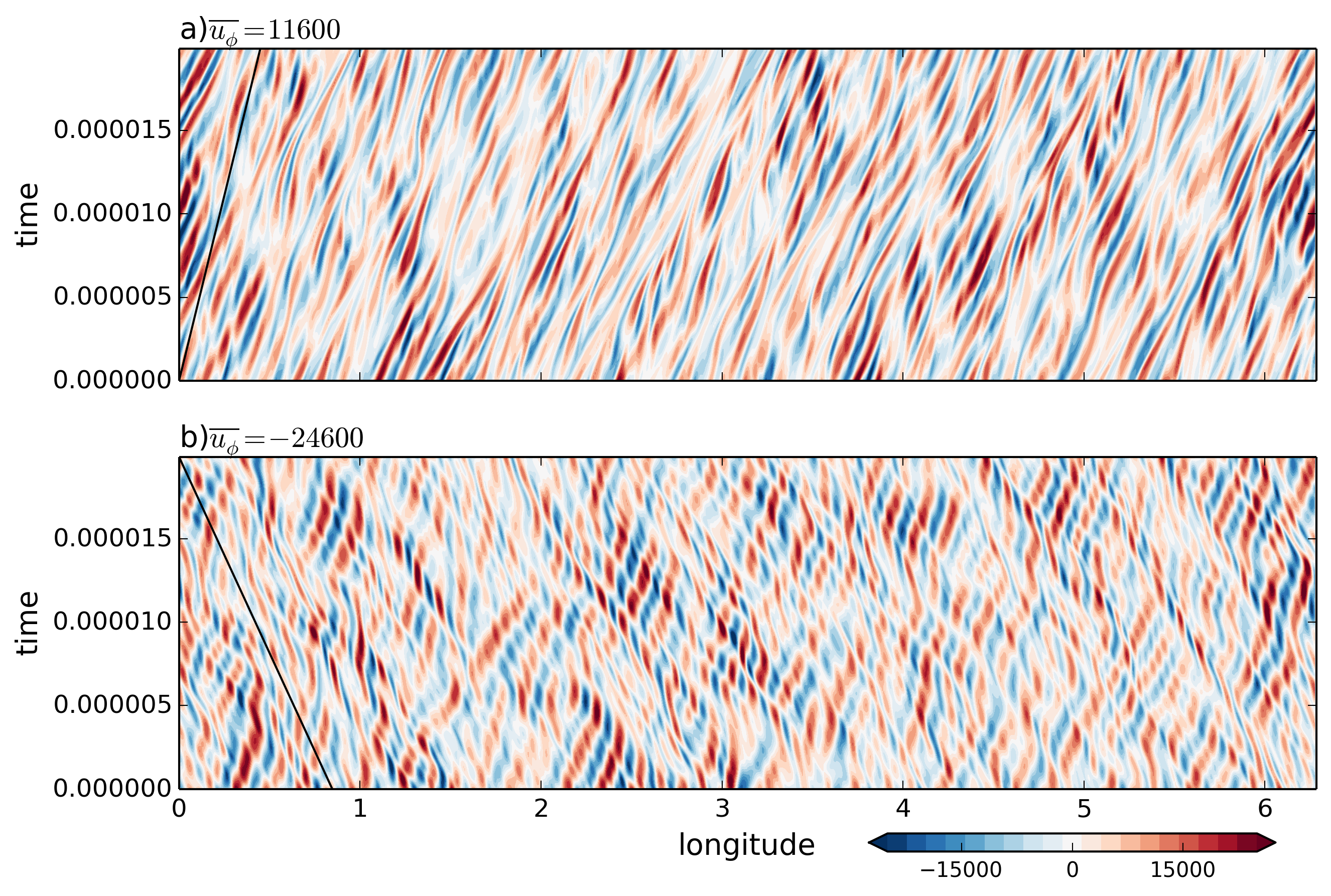}
   \caption{Hovm\"oller map (longitude-time) of the radial velocity $u_s$ at a fixed radius: (a) $s=0.51$ in a prograde jet ($\zw=11600$) and (b) $s=0.58$ in a retrograde jet ($\zw=-24600$) for $\Ra/\Ra_c=6.44$ ($\Ek=10^{-8}$, $\Pran=10^{-1}$). 
   The solid black line shows the drift due to the advection by the zonal velocity.
   \revision{The time interval corresponds to $2000$ rotation periods and $4$ convective turnover timescales (as defined in table~\ref{tab:list}).}}
   \label{fig:drift_eddies}
\end{figure*}

We now discuss the mechanism of formation of persistent zonal flows
based on the extensive literature on the subject, particularly in the context of the ocean and atmosphere dynamics \citep[\eg][]{Vallis2006}.
In our simulations, retrograde zonal flows are faster and sharper than the rounded prograde zonal flows. 
This asymmetry is indicative of an important dynamical difference between the two types of zonal jets, related to their formation mechanism and to the sign of $\beta$.
The asymmetry is also observed in 3D models of rotating spherical convection 
\citep[\eg][]{Heimpel2007,Gastine2014}
\revision{and in QG models with constant $\beta$ \citep[\eg][]{Tee12}.}
In the 3D simulations, the multiple zonal jets emerge inside the tangent cylinder, where $\beta$ is positive, 
so the prograde flows are faster and sharper than the retrograde flows. 

In quasi-geostrophic flows, the zonal flows are directly related to the distribution of potential vorticity (PV) \citep[\eg][]{Mcintyre2003}. 
In our system, the equation for the evolution of the potential vorticity, $\PV$, is 
\begin{equation}
	\pdt{\PV} + \vect{u} \cdot \nabla \PV = D + F ,
\end{equation}
where
\begin{equation}
	\PV=\frac{\vorz + 2\Ek^{-1}}{H},
\end{equation} 
$D$ represents the dissipation terms from the boundary and the bulk and $F$ the buoyancy source. $\vorz$ includes the axisymmetric and
non-axisymmetric components of the vorticity. 
In the absence of buoyancy sources and dissipative effects, $\PV$ is a materially invariant scalar so
it tends to be locally homogenised by the turbulence.  The zonal flows appear as a consequence of the mixing of PV with
\begin{equation}
	\overline{u_{\phi}}(s) = \frac{1}{s} \int_0^s s^{\prime} \overline{\vorz} ds' 
					= \frac{1}{s} \int_0^s s^{\prime} \pl H \overline{\PV} - 2 \Ek^{-1} \pr ds' .
\label{eq:uPV}
\end{equation}
In a sphere (where $H$ decreases with $s$, \ie $\beta<0$), regions of weak PV gradients correspond 
to prograde zonal jets, while regions of strong PV gradient correspond to retrograde zonal jets.
The PV distribution controls the quasi-geostrophic dynamics and, in particular, the propagation of Rossby waves. 
The equation for linear Rossby waves is 
\begin{equation}
	\frac{\partial \vorz'}{\partial t} + u_s H \frac{d \moyp{\PV}}{ds} = 0,
\end{equation}
where $\vorz'=\vorz-\moyp{\vorz}$ and 
\begin{equation}
	\frac{d\moyp{\PV}}{ds} = \frac{d\moyp{\vorz}/H}{ds} - \frac{2\beta}{\Ek H}.
\end{equation} 
The restoring mechanism of the Rossby waves is therefore stronger in the retrograde jets than in the prograde jets.
A strong mixing leads to the formation of a staircase of PV, \ie a succession of regions of homogeneous PV separated by regions of steep PV gradients \citep[\eg][]{Scott2012}. 
This occurs because perturbations to strong PV gradients are radiated as Rossby waves, an effect referred to as Rossby wave elasticity \citep{Mcintyre2008}.
Perturbations are therefore inhibited in the regions of strong PV gradients, while they are intensified in regions of weak PV gradients. 
This leads to a feedback mechanism that further steepens the gradients of PV \citep{Dri08}.
In a sphere, the relation between PV and zonal flows (\ref{eq:uPV}) implies that wide regions of weak PV gradients appear as large rounded prograde jets,
whereas narrow regions of steep PV gradients appear as sharp retrograde jets.

The radial profile of the axisymmetric potential vorticity is plotted in figure~\ref{fig:pv_staircase} for $\Ra/\Ra_c=9.02$.
In the range $s\in[0,0.8]$, the succession of weak and steep PV gradient regions is visible and forms a relatively mild PV staircase for this value of $\Ra$. 
Larger $\Ra$ are required to obtain well mixed PV regions and a better formed staircase profile.
The asymmetry of the jets is clearly visible in our simulations but retrograde jets are not particularly narrower than the prograde jets.
This is because the chosen delimitation of the jets ($\zw=0$) is, somewhat arbitrarily, defined with respect to the planetary rotation. The profile of 
$\moyp{q}$ shows that the regions of steep PV gradients (which correspond to the cores of the retrograde jets) are in fact much narrower than the regions of weak PV gradients. 
Another characteristic of the zonal flow is the robustness of the prograde jet at the centre in all our simulations: 
this is well explained by the mixing in the central region that leads to a local 
increase of $\PV$.
The PV mixing mechanism therefore provides a good explanation for the most notable features of the zonal flows observed in our simulations. 
Nevertheless, we note that alternative -- although not mutually exclusive -- mechanisms for the formation of persistent zonal flows have been put 
forward in the literature, such as resonant triad interactions \citep[\eg][]{Ped87}.

The PV distribution indicates that the non-axisymmetric dynamics inside the retrograde jets might be dominated by Rossby waves.
Assuming for simplicity $\vorz'=-\nabla_e^2\psi$ (see equation~\ref{eq:defPsi}), the dispersion relation of the Rossby waves is
\begin{equation}
	\omega = H \frac{d\moyp{\PV}}{ds} \frac{k_{\phi}}{|\vect{k}|^2},
\end{equation}
where $\omega$ is the frequency of the wave and $\vect{k}=(k_s,k_{\phi})$ is the wavenumber vector. 
Consequently, the azimuthal phase velocity, $v_p$, and azimuthal group velocity, $v_g$, of the Rossby waves are
\begin{equation}
	v_p = H \frac{d\moyp{\PV}}{ds} \frac{1}{|\vect{k}|^2}, \quad v_g = H \frac{d\moyp{\PV}}{ds} \frac{k_s^2-k_{\phi}^2}{|\vect{k}|^4}.
	\label{eq:velocityRW}
\end{equation}
In our system, the gradient of $\moyp{\PV}$ is positive so the Rossby waves always have a positive azimuthal phase speed.
To determine whether Rossby waves are present in the retrograde jets, we can track the direction of the azimuthal drift of the velocity patterns. 
Figure~\ref{fig:drift_eddies} shows the Hovm\"oller map (longitude-time)
of the radial velocity $u_s$ at a fixed radius in a prograde jet ($\zw=11600$) and in a retrograde jet ($\zw=-24600$). 
The solid black line represents the azimuthal drift due to the advection 
by the zonal velocity at this radius, $\Delta t= s\Delta \phi/\zw$. 
In the prograde jet, the radial velocity structures move in the prograde direction at a rate that is consistent with the advection by the zonal velocity,
and even faster for some structures. 
In the retrograde jet, the radial velocity patterns mainly move in the prograde direction. 
These patterns must therefore correspond to Rossby waves. The amplitude of the patterns of $u_s$ is modulated as they move in the prograde direction.
These modulations appear on neighbouring patterns and seem to travel in the retrograde direction. 
This observation is consistent with Rossby waves for which $k_s<k_{\phi}$ so their azimuthal group velocity is negative. 
In the frame of reference \revision{(\ie rotating at the rotation rate $\Omega$)}, the Rossby waves move at a velocity $v_p+\zw$. From figure~\ref{fig:drift_eddies}b,
we estimate that this velocity is approximately $20000$, which gives $v_p\approx45000$. By using the local value of the gradient of $\moyp{\PV}$ at this radius, 
we find that $|\vect{k}|^2\approx84$ from eq.~(\ref{eq:velocityRW}). From figure~\ref{fig:drift_eddies}b
we estimate that the azimuthal wavenumber is $k_{\phi}=m/s\approx 70$, and so, we deduce that $k_s\approx46$. 
This result is consistent with a negative azimuthal group velocity.

We might expect that the dynamical difference between prograde and retrograde jets
affects the temperature distribution \revision{because the Rossby waves might modify the transport properties of the flow}. We study this \revision{problem} in the next section.

\section{Effect of the zonal flows on the heat transport}
\label{sec:barrier}

\begin{figure*}
\centering
    \includegraphics[clip=true,width=14cm]{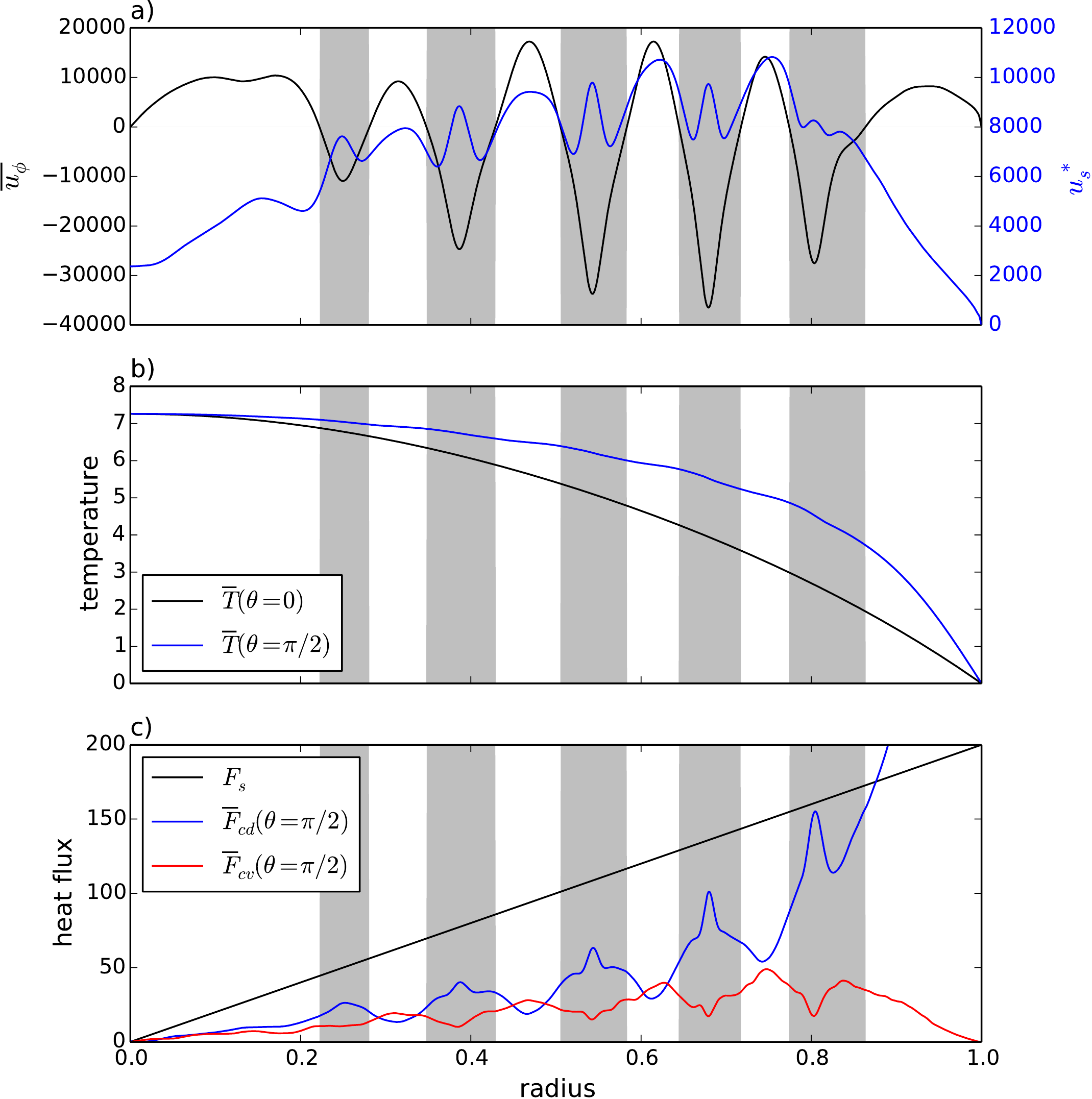}
   \caption{
   Radial profile of (a) the zonal velocity (black line) and the r.m.s radial velocity (blue), 
   (b) the axisymmetric temperature along the rotation axis (black) and in the equatorial plane (blue), 
   (c) the static heat flux (black), the conductive heat flux in the equatorial plane (blue) and the
   convective heat flux in the equatorial plane (red). All the quantities have been time-averaged.
   The grey bands correspond to the regions where the zonal flow is retrograde.
   The parameters are $\Ra/\Ra_c=9.02$, $\Ek=10^{-8}$ and $\Pran=10^{-1}$.}
   \label{fig:barrier}
\end{figure*}

\begin{figure*}
\centering
    \includegraphics[clip=true,width=14cm]{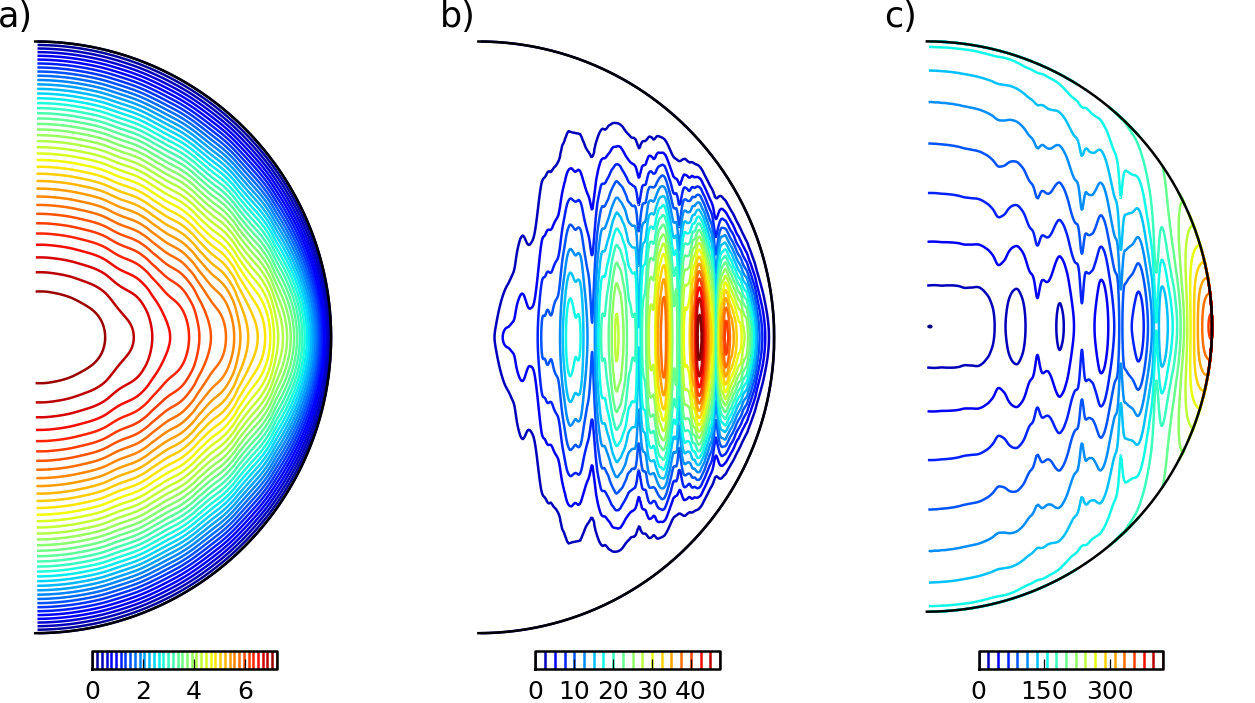}
   \caption{
   Meridional cross-section of the axisymmetric average of (a) the temperature $\moyp{T}=\moyp{\Theta}+T_s$, (b) the convective heat flux $\moyp{F}_{cv}$,
   and (c) the conductive heat flux $\moyp{F}_{cd}$. All the quantities have been time-averaged. Same parameters as figure~\ref{fig:barrier}.}
   \label{fig:Tmer_Ra7e11}
\end{figure*}

\begin{figure*}
\centering
    \includegraphics[clip=true,width=14cm]{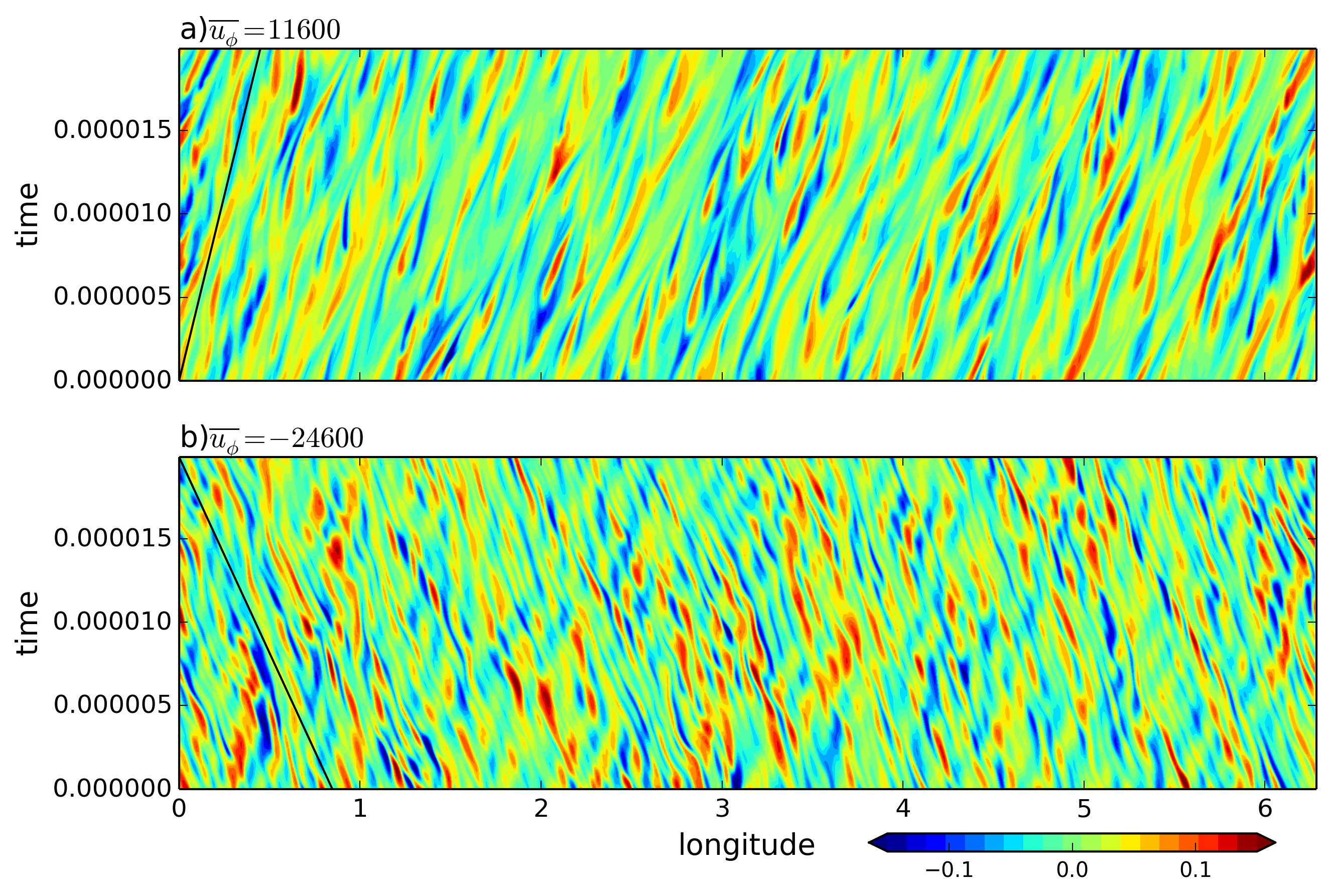}
   \caption{Hovm\"oller map of the non-axisymmetric temperature perturbation, $\Theta'$, at a fixed radius: (a) $s=0.51$ in a prograde jet and (b) $s=0.58$ in a retrograde jet
   for the same parameters as figure~\ref{fig:drift_eddies}. The solid black line shows the drift due to the advection by the zonal velocity.
    \revision{The time interval corresponds to $2000$ rotation periods and $4$ convective turnover timescales (as defined in table~\ref{tab:list}).}}
   \label{fig:drift_T}
\end{figure*}

First, we can assess the effect of the zonal flow on the amplitude of the radial flow by comparing the profile of the zonal velocity
to the radial profile of the r.m.s radial velocity, $u_s^{\ast}$, which is calculated as
\begin{equation}
	\us(s) = \frac{1}{\Delta t} \int_{\Delta t} \pl \frac{1}{2\pi s} \int_0^{2\pi} u_{s}^2(s,\phi,t) s d\phi \pr^{1/2} dt.
\end{equation}

Figure~\ref{fig:barrier}a shows these profiles for $\Ra/\Ra_c=9.02$.
The zonal flow (black line) is plotted according to the left axis and  $u_s^{\ast}$ (blue line) is plotted according
to the right axis. Both profiles are time-averaged. 
In the inner convective region ($s\lesssim 0.8$), the local maxima of $u_s^{\ast}$ correlates well with 
the extrema of $\zw$ (the cores of the jets), \ie the zeros of the radial shear $|\partial_s \zw|$, 
while the local minima of $u_s^{\ast}$ correlates with maxima of  $|\partial_s \zw|$ (the flanks of the jets).
The reduction of $u_s^{\ast}$ in the flank of a jet can reach 30\% of its value in 
the core of the neighbouring jets. The radial velocity is therefore impeded by the radial shear and maximised in the cores of jets, irrespective of their sign.
Thus the dynamical difference between prograde and retrograde jets cannot be directly diagnosed on this profile.

To determine how the temperature field is affected by the zonal flows, figure~\ref{fig:Tmer_Ra7e11}a shows a
meridional slice of the 
axisymmetric temperature, $\moyp{T}=\moyp{\Theta}+T_s$, averaged in time for the same simulation.
The isotherms have an ellipsoidal shape which is elongated towards the equator.
To complement this figure, the radial profiles of the axisymmetric temperature along the rotation axis, $\moyp{T}(\theta=0)$, and in the equatorial plane, $\moyp{T}(\theta=\pi/2)$, are
plotted in figure~\ref{fig:barrier}b. In the equatorial plane, the temperature has a flatter profile than along the rotation axis for $s<0.8$, which explains the ellipsoidal shape 
of the isotherms. 
The thermal boundary layer is more pronounced in the equatorial plane. \revision{Nevertheless} it remains thick (much thicker than the Ekman layer)
because $\Pran<1$ and the Rayleigh number is moderate.
The disparity between the temperature along the axial and equatorial directions is due to preferential direction of the convective flows in rapidly rotating convection.
This effect is also observed in 3D models \citep[\eg][]{Zhang1991,Yad16}, but is amplified here by the use of a quasi-geostrophic model. 

On top of their ellipsoidal shape, the isotherms also have undulations of small amplitude. These undulations are located at the same distance from the rotation axis for each isotherm
so they are likely due to the presence of zonal flows.
This causal link is visible in figure~\ref{fig:barrier}b where we use the radial profile of $\moyp{T}(\theta=\pi/2)$
as a proxy for the $z$-averaged axisymmetric temperature.  
The temperature profile is relatively flat in the prograde zonal jets. By contrast it is significantly steeper in the core of the retrograde jets. 
This indicates that the mean temperature is affected by the PV distribution: in the core of the retrograde jets, 
the inhibition of the turbulence and the dominance of the Rossby waves imply that the temperature cannot be efficiently homogenised.

This effect can be quantified by measuring the heat fluxes carried by convection and conduction through the system. 
In the steady state, the volume average of the heat equation implies that
\begin{equation}
	\frac{1}{\mathcal{S}} \int_{\mathcal{S}} \pl F_{cv} + F_{cd} \pr d\mathcal{S} = F_s,
\end{equation}
where the convective heat flux is
\begin{equation}
	F_{cv} = \Theta u_r,
\end{equation}
the conductive heat flux,
\begin{equation}
	F_{cd} = - \frac{1}{\Pran} \frac{\partial T}{\partial r},
\end{equation}
the static heat flux,
\begin{equation}
	F_s =  - \frac{1}{\Pran} \frac{dT_s}{dr} = \frac{2}{\Pran^2} r, 
\end{equation}
and $\mathcal{S}$ is a spherical surface. 
Figure~\ref{fig:Tmer_Ra7e11} shows the meridional slices of the axisymmetric averages of the convective heat flux, $\moyp{F}_{cv}$, and
of the conductive heat flux, $\moyp{F}_{cd}$.
The fluxes are time-averaged.
Both fluxes have a banded structure aligned with $z$. The convective heat flux is mainly concentrated around
the equatorial plane and is weak near the axis. The conductive heat flux is maximum near the equator.
For this simulation at $\Pran=10^{-1}$, the conduction carries a larger part of the heat than the convection in most of the domain,
despite the large values of the radial velocity (of the order of $10^4$). 

The thermal contrast between prograde and retrograde jets can be clearly observed in 
the profiles of the axisymmetric heat fluxes in the equatorial plane shown in figure~\ref{fig:barrier}c.
For comparison, the static heat flux $F_s$ is also shown. 
Note that the profile of $\moyp{F}_{cv}$ and $\moyp{F}_{cd}$ at a given latitude are not representative of the spherical averages so their sum
is not equal to $F_s$ at each radius.
The local maxima of the convective flux are located in the prograde zonal jets. 
The decrease of the convective flux matches the decrease of $u_s^{\ast}$ in the shear layers. However
the convective flux systematically reaches minima in the core of the retrograde jets while the 
radial velocity recovers there. 
The conductive flux is boosted in the shear layers, but even more so in the retrograde jets, where 
it is much larger than the convective flux. 
The balance between the thermal processes is very different in the prograde and retrograde jets: the prograde jets are regions where about half of the heat 
is carried by the convection and the temperature is fairly uniform, whereas
most of the heat is carried by the conductive flux in the retrograde jets.
This occurs despite high values of the r.m.s radial velocity in the retrograde jets, so the radial flow
must not be well correlated with the temperature perturbation in these regions.
Figure~\ref{fig:drift_T} show the Hovm\"oller map for the non-axisymmetric temperature perturbation, $\Theta'=\Theta-\moyp{\Theta}$,
for the same parameters and radius as figure~\ref{fig:drift_eddies}.  
In both prograde and retrograde jets, the temperature perturbation drifts in the direction of the zonal flow 
with a drift rate consistent with the advection by the zonal velocity.
\revision{In the core of the retrograde jets, which are dominated by the propagation of Rossby waves, the radial velocity and temperature perturbation are visibly not well correlated. 
The inefficiency of the Rossby waves at transporting heat outwards explains the weak convective heat flux there.}
In the prograde jet, $\Theta'$ is well correlated with $u_s$, which is consistent with the high convective heat flux.

In summary, our results show that the core of the retrograde jets -- and not only the regions of intense shear -- act as primary bottlenecks to the convective heat transport.

\section{Discussion}
\label{sec:discussion}

\begin{figure}
\centering
    \includegraphics[clip=true,width=8cm]{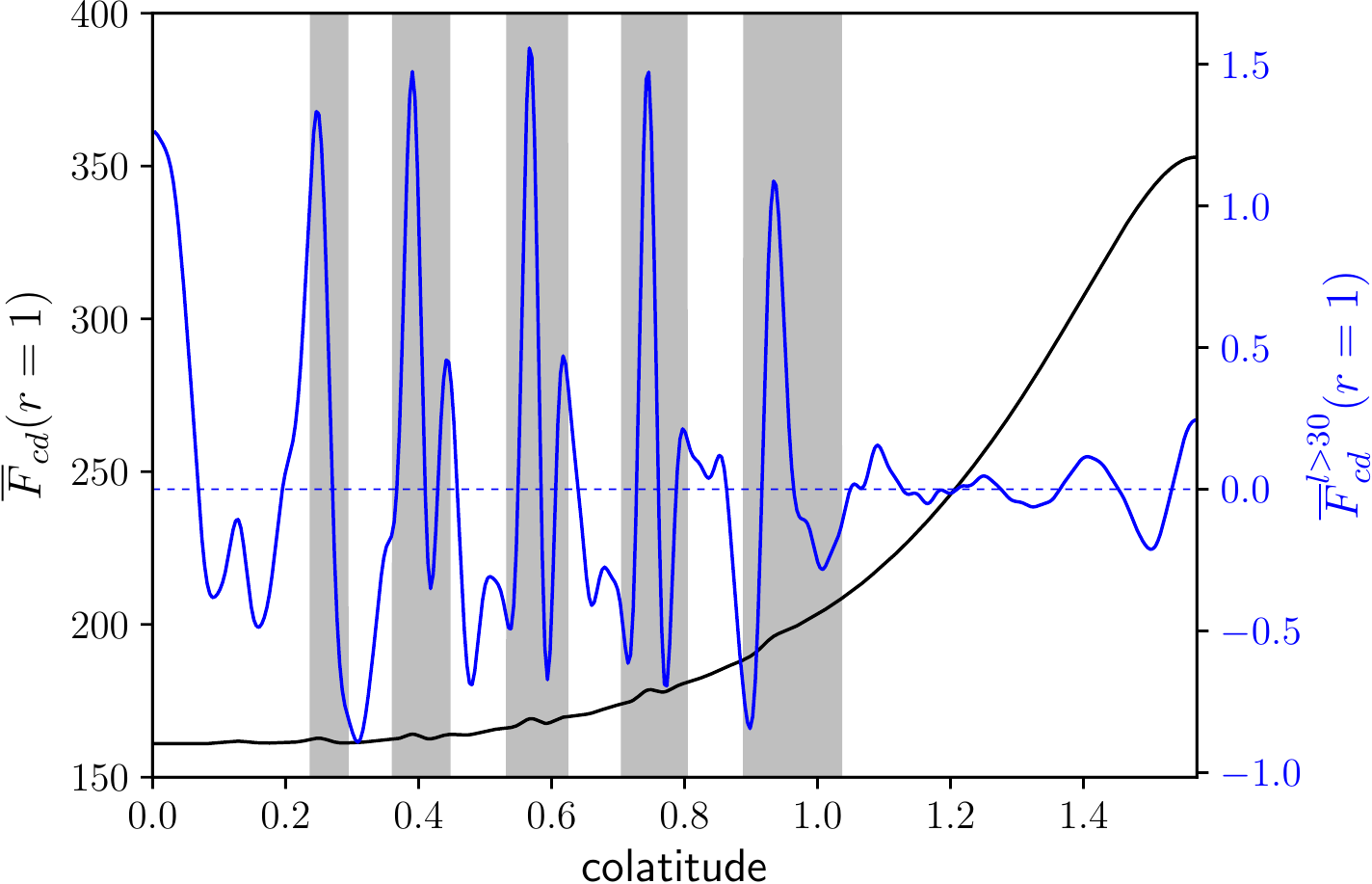}
   \caption{Latitudinal profile of the axisymmetric heat flux \revision{(black line)} 
   from the North pole to the equator at the outer boundary $r=1$ for $\Ek=10^{-8}$, $\Pran=10^{-1}$ and $\Ra/\Ra_c=9.02$.
   \revision{The blue line shows the axisymmetric heat flux where the spherical harmonics coefficient of degrees $l\leq30$ 
   have been filtered out and is plotted according to the right axis.}
   The grey bands correspond to the regions where the zonal flow is retrograde.}
   \label{fig:qsurf}
\end{figure}

In this paper, we have studied the convective structures and zonal flows that form in rotating thermal convection for values of the Prandtl number
relevant for liquid metals ($\Pran=\mathcal{O}(10^{-1})$) and low Ekman ($\Ek=\mathcal{O}(10^{-8})$) and Rossby numbers ($\Ro_c<10^{-2}$). 
In order to reach low values of the Ekman number, we have used a hybrid numerical model that couples a quasi-geostrophic 
approximation for the velocity to a 3D temperature field. Convection is driven by internal heating in a full sphere geometry. The model
includes Ekman pumping to mimic no-slip boundary conditions. 
We focus on the intense zonal flows that emerge on the strong branch of convection, which was described by \citet{Guervilly2016}
using the same hybrid QG-3D model and by \citet{Kaplan2017} in a fully 3D model.
Persistent multiple zonal jets of alternating sign form due to the mixing of potential vorticity and exert a strong feedback on the convection. 
An upscale energy transfer takes place and the integral convective lengthscale increases with the vigour of the convection.
The convective lengthscale and the zonal jet width are closely related:  the radial shear exerted by the zonal flow on the radial velocity limits the size of the convective eddies,
while the typical mixing length of the potential vorticity depends on the size of the most energetic convective eddies.
The convective lengthscale varies radially in agreement with the Rhines scale \citep{Rhi75}, \ie as the square root of $\beta$, where $\beta$ measures 
the slope of the boundaries. \revision{However the convective scale increases more slowly with the convective speed (following a power law of exponent $0.37$)
than predicted by the Rhines scale (power law of exponent $0.5$).}
 
In our quasi-geostrophic model, convection carries heat mostly in the direction perpendicular to the rotation axis.
We have shown that the principal barrier to this convective heat transport is located in the cores of the retrograde zonal jets.
This is due to the formation of a staircase of potential vorticity: steep and weak gradients of the potential vorticity correspond to retrograde and prograde zonal jets, respectively.
The steep PV gradients inhibit the eddies and favour the propagation of Rossby waves, which are inefficient at carrying heat outwards. 
The occurrence of eddy-transport barriers associated with strong zonal jets is well-documented in the atmospheric dynamics context \citep[][]{Dri08}.
In our simulations, these barriers lead to the steepening of the mean temperature gradient in the core of the retrograde jets, and there, the heat is largely carried by conduction. 
The unfavourable effect of the shear layer in the flanks of the zonal jets on the convective heat transport is secondary by comparison.  
To illustrate the thermal signature of the retrograde jets at the surface, we plot in
figure~\ref{fig:qsurf} the latitudinal profile of the axisymmetric heat flux at $r=1$. The most noticeable feature is that the heat flux is maximal at the equator, which is expected
as the convective transport is largely perpendicular to the rotation axis, \revision{and is also observed in 3D models \citep[\eg][]{Zhang1991,Yad16}.}
This enhanced heat transport in the equatorial regions compared with the polar regions is used in a number of models of the Earth's inner core growth 
to explain the observed seismic anisotropy \citep[\eg][]{Yos96,Deguen2009}. 
Of greater interest here are the more subtle variations of the surface heat flux at higher latitudes: the cores of the retrograde jets
are characterised by local maxima of the (conductive) surface heat flux. This occurs because the axisymmetric temperature gradient is steepest in these regions. 
\revision{To highlight the small-scale anomalies of the surface heat flux and quantify their amplitude, 
we filter out the coefficients of the spherical harmonics of degree $l$ smaller than 30. The filtered profile is plotted according to the right axis of figure~\ref{fig:qsurf}
in blue. It shows that the heat flux anomalies due to the presence of the zonal flows are sharp and of a small amplitude, approximately $1\%$ of the mean surface heat flux.}
The thermal signal associated with zonal jets provides useful information in the context of the gas giant planets because the power emitted at the surface 
of the planet can be measured.
For Jupiter and Saturn, the emitted power is approximately uniform in latitude with variations of small amplitude that resemble the structure of the zonal flows
at the surface \citep[][]{Pirraglia1984,Li2010}.

For increasing thermal driving (out of reach of our current computational resources), 
we expect the convective eddies to further enhance the mixing of potential vorticity in the prograde jets. The PV staircase would then become sharper with 
narrower regions of steep PV gradients, and hence narrower retrograde jets. 
\revision{This sharpening of the PV staircase in rapidly-rotating convection is observed by \citet{Verhoeven2014} at large Rayleigh numbers (20 times supercritical)
in 2D numerical simulations using the anelastic approximation. In their study, the $\beta$ effect is due to the compressibility of the fluid \citep{Ing82,Gla09}
and leads to the formation of multiple zonal jets, similarly to the topographic $\beta$ effect as studied here. 
They show that the sharpening of the PV staircase results in the sharpening of the entropy staircase (analogous to the temperature staircase in the Boussinesq approximation).}
In quasi-geostrophic systems with vigorous convection, we thus expect the heat transport process to be heterogeneous with wide convective regions 
separated by narrow conducting bands; the efficiency of the heat transfer throughout the whole system
would largely be controlled by the efficiency of the conducting process across the retrograde jets, and thus, by the width of
these conducting bands. This process is partly analogous to the occurence of layering in double-diffusive convection where the heat transfer is controlled
by the flux through the interface between overturning layers \citep{Turner1985}. 

\revision{The main features of the nonlinear quasi-geostrophic dynamics discussed in this paper do not crucially rely on the temperature being 3D. Consequently, 
we expect that QG models of rotating convection using a 2D temperature field would be able to reproduce our observations qualitatively. QG-2D models could be used to pursue 
this study at lower Ekman numbers and larger Rayleigh numbers.
The numerical framework used in the hybrid QG-3D model is however well suited to explore the possible existence of dynamos driven by quasi-geostrophic flows
at low magnetic Prandtl numbers \citep[\eg][]{Gil11}. The dynamo problem indeed requires to treat the magnetic field in 3D.  
We shall investigate QG dynamos in a forthcoming study.}  

\section*{Acknowledgements}
CG was supported by the Natural Environment Research Council under grant NE/M017893/1.
PC acknowledges the Agence Nationale de la Recherche for supporting this project under grant ANR TuDy.
This work was undertaken
on TOPSY of the HPC facilities at Newcastle University and on the facilities of N8 HPC Centre of Excellence, provided and 
funded by the N8 consortium and EPSRC (Grant EP/K000225/1) and co-ordinated by the Universities of Leeds and Manchester. 
We are grateful to the referees for helpful comments that improved the manuscript.

\end{document}